\documentclass[useAMS,usenatbib,a4]{mn2e}
\usepackage[english]{babel}
\usepackage[utf8]{inputenc}
\usepackage{makeidx}
\usepackage{array,multirow}
\usepackage{rotating}
\usepackage{color}
\usepackage{epsfig}
\usepackage{epstopdf}
\usepackage{float}
\restylefloat{table}
\usepackage{here}
\usepackage{fixltx2e} 
\usepackage[font=small,format=plain,labelfont=bf,up,textfont=it,up]{caption}
\usepackage{graphicx}
\usepackage{verbatim}
\usepackage{amsmath}
\usepackage{multicol}
\usepackage{subcaption}
\usepackage{xcolor}
\usepackage[colorlinks=false, linkbordercolor=blue,draft]{hyperref}
\DeclareGraphicsExtensions{.pdf,.png,.jpg,.mps}

\usepackage{xspace}
\def\ap{Alcock-Paczy\'{n}ski test\xspace}

\newcommand{\hmpc}{$\mathrm{h^{-1}Mpc}$\xspace}

\newcommand{\vide}{\tt VIDE\normalfont\xspace}
\newcommand{\zobov}{\tt ZOBOV\normalfont\xspace}

\renewcommand{\tabcolsep}{-1cm}

\title[Mastering peculiar velocities in cosmic voids]{Mastering the effects of peculiar velocities in cosmic voids}
\author[A. Pisani et al.]{Alice Pisani$^{1,2,3}$\thanks{E-mail:
pisani@cppm.in2p3.fr (AP)
}, P. M. Sutter$^{4,5,7}$ and B. D. Wandelt$^{2,3,6}$
\\ \\
$^{1}$Centre de Physique des Particules de Marseille, Aix-Marseille Universit\'e, CNRS/IN2P3, Marseille, France\\
$^{2}$Sorbonne Universit\'es, UPMC (Paris 06), UMR7095, Institut d’Astrophysique de Paris, 98bis Bd. Arago, F-75014, Paris,
France\\
$^{3}$CNRS, UMR7095, Institut d’Astrophysique de Paris, 98bis Bd. Arago, F-75014, Paris, France\\
$^{4}$INFN - National Institute for Nuclear Physics, via Valerio 2, I-34127 Trieste, Italy\\
$^{5}$INAF - Osservatorio Astronomico di Trieste, via Tiepolo 11, 1-34143, Trieste, Italy\\
$^{6}$Departments of Physics and Astronomy, University of Illinois at Urbana-Champaign, Urbana, IL 61801, USA\\
$^{7}$Center for Cosmology and Astro-Particle Physics, Ohio State University, Columbus, OH 43210}

\begin{document}

\date{Accepted ----. Received ---; in original form ---}

\pagerange{\pageref{firstpage}--\pageref{lastpage}} \pubyear{2015}

\maketitle

\label{firstpage}

\begin{abstract}

How do peculiar velocities affect observed voids? To answer this question we use the VIDE toolkit to identify voids in mock galaxy populations embedded within an N-body simulation both with and without peculiar velocities included. We compare the resulting void populations to assess the impact on void properties. We find that void abundances and spherically-averaged radial density profiles are mildly affected by peculiar velocities. However, peculiar velocities can distort by up to 10\% the shapes for a particular subset of voids depending on the void size and density contrast, which can lead to increased variance in \ap.  We offer guidelines for performing optimal cuts on the void catalogue to reduce this variance by removing the most severely affected voids while preserving the unaffected ones. In addition, since this shape distortion is largely limited to the line of sight, we show that the void radii are only affected at the $\sim$ 10\% level and the macrocenter positions at the $\sim$ 20\% (even before performing cuts), meaning that cosmological probes based on the Integrated Sachs-Wolfe and gravitational lensing are not severely impacted by peculiar velocities.

\end{abstract}


\section{Introduction}

In the era of the large scale surveys, extracting cosmological information from the large scale structure of the Universe is a major challenge. The distribution of galaxies in the Universe follows a cluster-sheet-void-filament pattern called the cosmic web \citep{Bond1996}. Recently, the underdense regions of the cosmic web (discovered in 1978, see \cite{Gregory1978}) captured the interest of cosmologists, as they open up new avenues of constraining models of the evolution of the Universe. 
Cosmic voids promise to be dynamically simpler than overdense structures \citep{Weygaert1993,Hamaus2014} and fill most of the Universe \citep{Fairall1998,Hoyle2002,Bos2012a}: they are potentially powerful tools to constrain cosmological models. 

Voids are being used for a wide range of applications, spanning from their use as standard spheres through the \ap \citep{Sutter2012b,Sutter2014}, to the study of the late-time Integrated Sachs-Wolfe effect in $f(R)$ models \citep{Cai2014}, or to constrain models of modified gravity \citep{Li2011, Spolyar2013, Clampitt2013, Zivick2014, Cai2014a}, models of coupled dark energy \citep{Sutter2014b} and cosmological models of massive or massless neutrinos \citep{Massara2015}.

In this context, it is crucial to properly model the impact of systematic effects, which might undermine the quality of cosmological constraints from cosmic voids. 
Voids probe a different dynamical regime of structure formation than high density regions, they show a quasi-linear behaviour \citep{Weygaert1993, Hamaus2014} and are surrounded by mildly non-linear walls \citep{Paz2013}. For this reason, they present different systematics than cosmological tools from high-density regions. 

The peculiar velocities of galaxies are the main systematic when using voids for cosmology (as discussed by e.~g.~ \cite{Ryden1995} and \cite{Sutter2014}). 
The position of each galaxy is measured in redshift space, thus the peculiar velocity of the galaxy changes the galaxy position by adding a component to the distortion due to the expansion of the Universe --- hence affecting our measures of the cosmic web. It is therefore important to assess the impact of such velocities on voids when aiming to a cleaner and optimal extraction of the cosmological signal. 
Recently \cite{Lavaux2012} studied the statistical impact of velocities on a dark matter simulation, and found a uniform bias in the Alcock-Paczy\'{n}ski measurement when peculiar velocities were included. This was confirmed by \cite{Sutter2014} for mock void populations, but those works did not examine in detail the source of that bias.

Other works focused on avoiding such bias, for example \cite{Pisani2014} presented a method to extract information from voids in a model-independent way through the reconstruction of voids density profiles in real space, that promises to improve the modelling of systematics for future surveys. Also, \cite{Hamaus2014b} introduced void auto-correlations as a promising statistic with reduced systematics impact. 
Nevertheless in most cosmological uses of voids, we lack of a detailed analysis of the velocity effects. While a full treatment of such effects on voids could lead to a reduction of systematics, its implementation is non-trivial \citep{Paz2013}.

Thus, in this work, we focus on the effect of velocities on void statistics and properties using mock galaxy catalogues, to make contact with voids measured with current surveys and used to constrain cosmology (as in \cite{Sutter2014}). The analysis of such effects is instrumental to a greater understanding of void systematics and avenues for correcting for their impacts. Rather than only examining a single void in isolation we study an ensemble of voids evolved in a full simulation, analysing their properties both on a one-to-one basis and on average. 

Works based on cosmological analysis of voids from real data (\textit{e.~g.}~ISW \citep{Cai2014}, lensing \citep{Melchior2014}, \ap \citep{Sutter2014}, abundances \citep{Pisani2015}, which could be affected by peculiar velocities, can use our results as a measure of the dynamical effects affecting the observed voids, and consider the guidelines emerging from this work to exclude the most affected voids from the analysis. Using such guidelines, the signal-to-noise ratio for cosmological measurements such as the \ap \citep{Alcock1979} promises to be enhanced.

The paper is organised as following: in Section 2 we present the void finder and simulation employed for this work, as well as the matching algorithm and methodology used to compare the two void catalogues (with and without velocities). In Section 3 we present the results: we discuss which voids survive after the application of peculiar velocities and analyse their features. In Section 4 we focus on one-to-one void comparisons to measure the impact of peculiar velocities on void properties, such as relative ellipticities, radii, and macrocenter positions.
We then analyse the effect on void stacks, widely used for the \ap and for applications such as the integrated Sachs-Wolfe effect (ISW) or lensing; and on abundances. Finally, in Section 5 we conclude with a discussion on how to reduce the impact of systematics related to peculiar velocities, we comment the results and their applicability for current and future surveys.

\section{Analysis method}
\subsection{Void finder}

The void finder used for this work is \vide, an improved version of \zobov \citep{Neyrinck2008}. The \vide toolkit \citep{Sutter2014a} finds cells in a distribution of tracers by means of the Voronoi tessellation and defines basins joining cells through the watershed transform. 
Following \zobov 's methodology, \vide does not merge zones if the minimum density along the ridge between zones is higher than 0.2 times the mean density of the simulation. As illustrated by \cite{Neyrinck2008}, this criterion prevents voids from growing into haloes. \vide provides the full void hierarchy, but for this work we will consider only parent voids with central densities less than $0.2 \bar{\rho}$ (see \cite{Sutter2014a} for details). Other choices may also be considered (see, for example, \cite{Nadathur2014}); according to the discussion in \cite{Lavaux2012} and \cite{Sutter2012b}, we focus on this choice to measure the impact of velocities in the framework of recent applications of the \ap.

With the convention used by \vide, the total volume $V$ of the void is the volume of all the Voronoi cells composing the void. The effective radius $R_{\rm eff}$ of the void can thus be defined through the volume as the radius of a sphere with volume $V$:
\begin{equation}
R_{\rm{eff}}\equiv\bigg( \frac{3}{4\pi}V\bigg)^{1/3}
\end{equation}
We also define the void macrocenter: it is the volume-weighted center of the void Voronoi cells. We remind that \vide naturally excludes voids with effective radius below the mean particle separation. This allows to set a limit for voids of too small size, affected by shot noise. 
Finally, we recall that \vide computes the shape of a void by taking the $N_{\rm{p}}$ void member particles and constructing the inertia tensor:

\begin{gather}
M_{xx}=\sum_{i=1}^{N_{\rm{p}}}(x_{i}^{2}+z_{i}^{2})\\ \nonumber
M_{xy}=-\sum_{i=1}^{N_{\rm{p}}}x_{i}y_{i}
\end{gather}

where $x_{i}$, $y_{i}$, and $z_{i}$ are the coordinates of the particle $i$ relative to the void macrocenter. Defining $J_{1}$ and $J_{3}$ as the smallest and largest eigenvalues of the inertia tensor, we obtain the definition of the ellipticity of the void:

\begin{equation}
\epsilon=1-\bigg(\frac{J_{1}}{J_{3}}\bigg)^{1/4}
\end{equation}

The macrocenter, the ellipticity and the effective radii of voids allow to analyse the properties of voids. 

\subsection{Simulation and HOD details}

The simulation we use in this work is a 1 $\mathrm{h^{-1}Gpc}$ box size dark matter N-body simulation, for which accuracy and error behaviour have been improved using the 2HOT code \citep{Warren2013} for cosmological volumes. It contains $1024^{3}$ particles and has a particle resolution of $7.36 \times 10^{11} h^{-1} M _{\odot}$. We used {\tt 2LPTIC} \citep{Crocce2006} and {\tt CLASS}\citep{Blas2011} to generate initial conditions. The {\tt 2HOT} code operation scales as $N{\rm log}N$ in the number of particles. More details on the simulation can be found in \cite{Sutter2014c}. 

To obtain two mock catalogues, we apply the {\tt Rockstar} halo finder and use it as an input for an Halo Occupation Distribution model \citep{Tinker2006, Zheng2007}. The model assigns to each dark matter halo of mass $M$ a central galaxy and satellite galaxies, the mean number of central galaxies and satellites is described by:

\begin{gather}
\big \langle N_{\rm{cen}}(M)\big \rangle =\frac{1}{2}\bigg[ 1+\rm{erf}\bigg(\frac{\rm{log} \it M-\rm{log} \it M_{\mathrm{min}}}{\sigma_{\rm{log} \it M}}\bigg)\bigg ]\\ \nonumber
\big \langle N_{\rm{sat}}(M)\big \rangle =\big \langle N_{\rm{cen}}(M)\big \rangle \bigg(\frac{M-M_{0}}{M^{'}_{1}}\bigg)^{\alpha}
\end{gather}
where we have $\sigma_{\rm{log} \it M}$, $M_{\rm min}$, $M_{0}$, $M_{1}^{'}$ and $\alpha$ as free parameters which are set to match the properties of a given galaxy population (see \cite{Sutter2014c}). 

We produce two galaxy catalogues, to mimic a high- and a low-resolution galaxy sample (with a tracer density of $2\times10^{-3}$ per cubic \hmpc for the \textit{HighRes} sample and $3\times10^{-4}$ per cubic \hmpc for the \textit{LowRes} sample). We assign the halo velocity for each galaxy, and additionally give a random peculiar velocity to each satellite galaxy by drawing from a Maxwellian distribution with mean equal to the dark matter halo velocity dispersion. We then use these galaxy samples to study the effects of peculiar velocities on the finding of voids in a realistic situation by comparing mocks with and without the presence of velocities in the high and low density case.

\renewcommand{\tabcolsep}{0cm}
\begin{figure*}
\begin{tabular}{ccc}
  \includegraphics[width=0.7\columnwidth, angle=0]{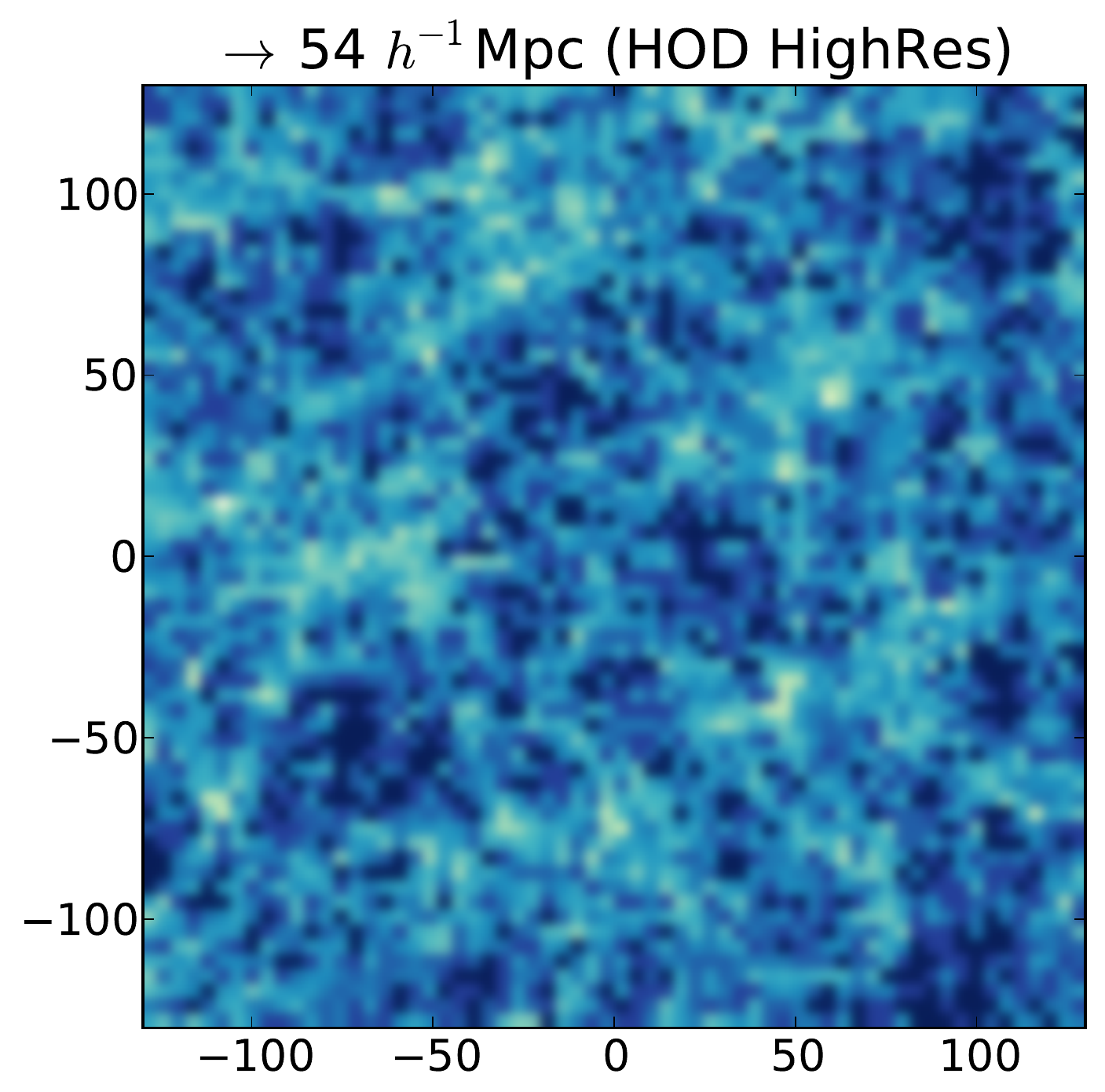}&
\includegraphics[width=0.7\columnwidth, angle=0]{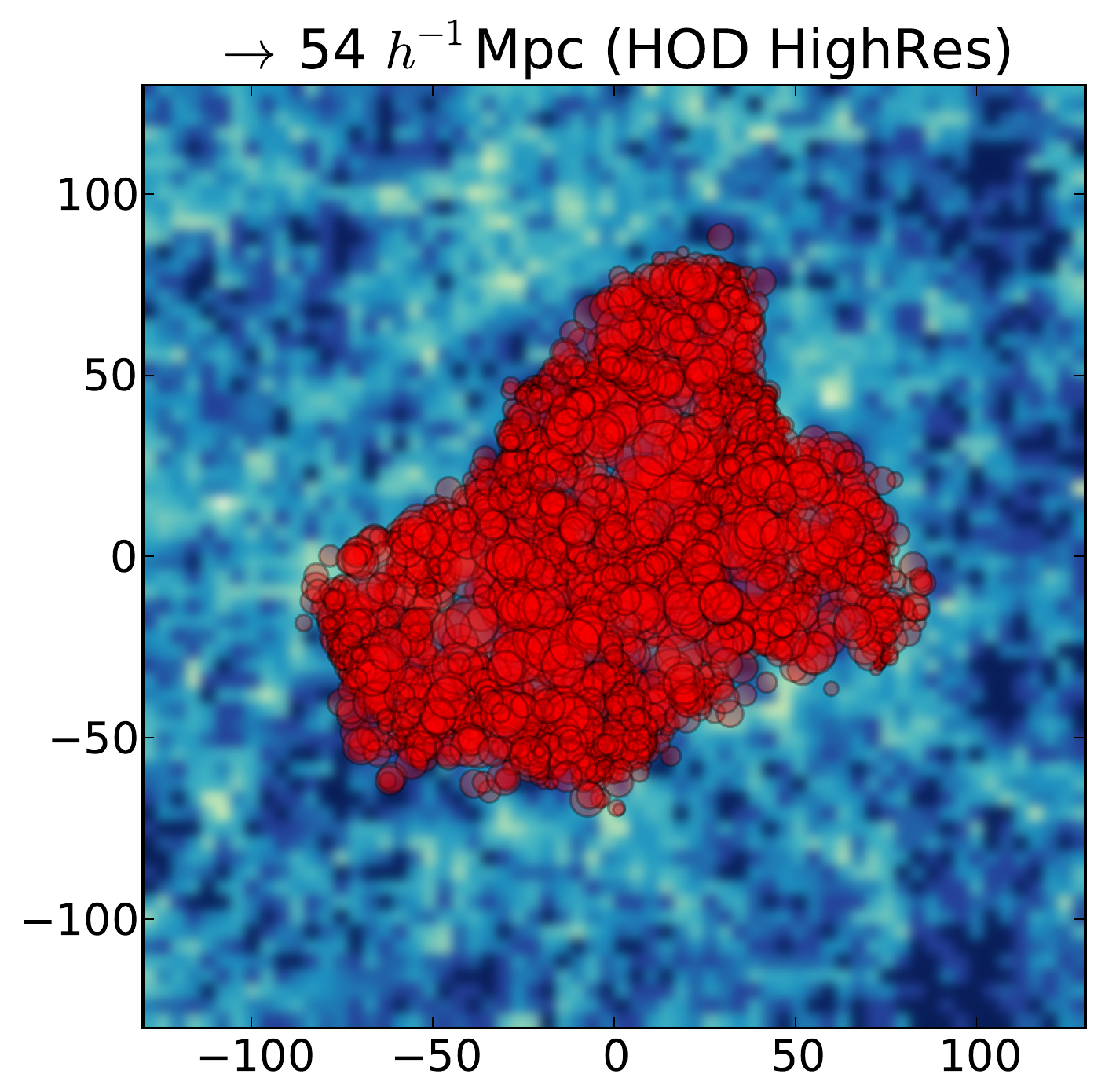}&
    \includegraphics[width=0.7\columnwidth, angle=0]{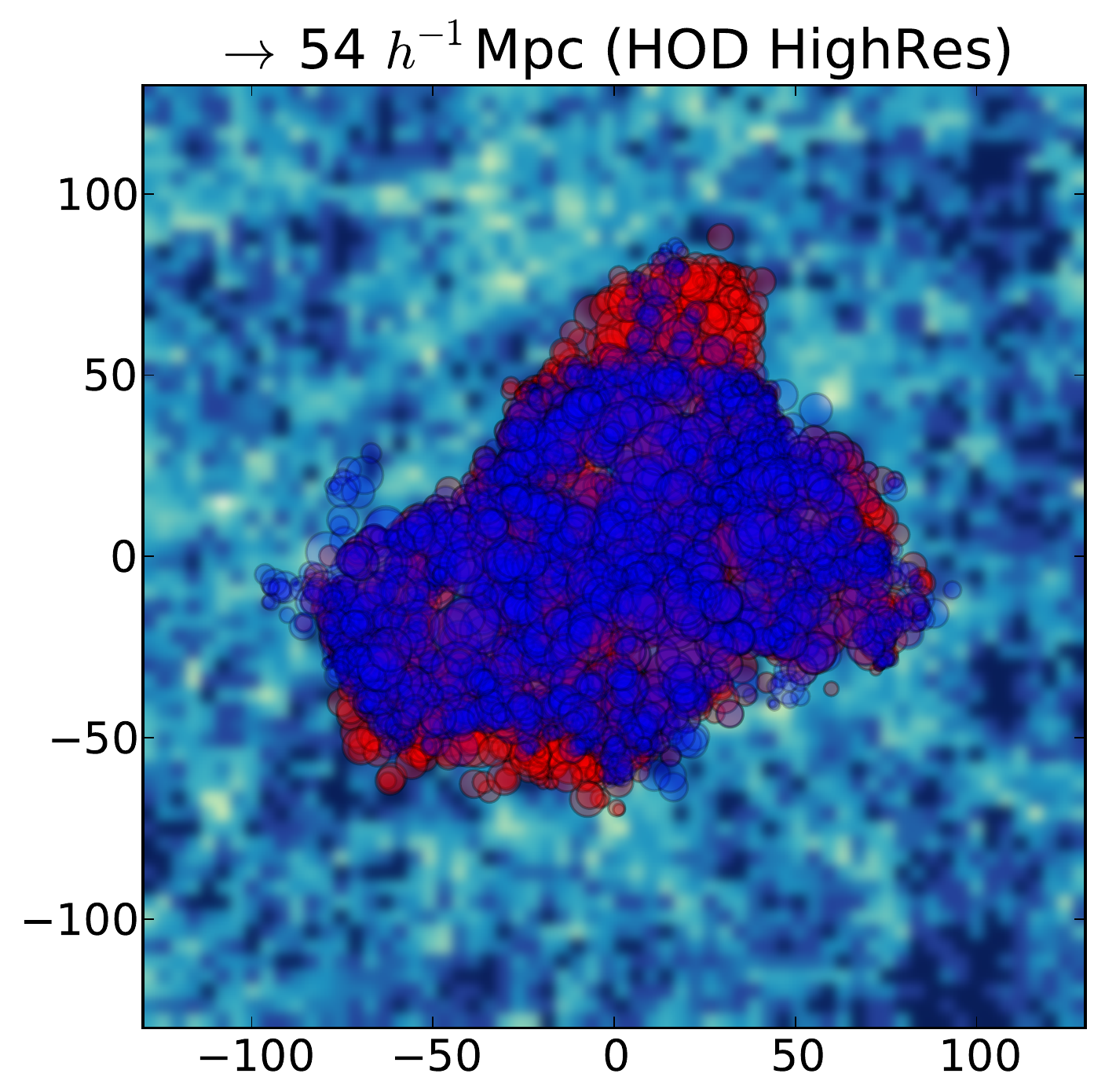}\\
      \includegraphics[width=0.7\columnwidth, angle=0]{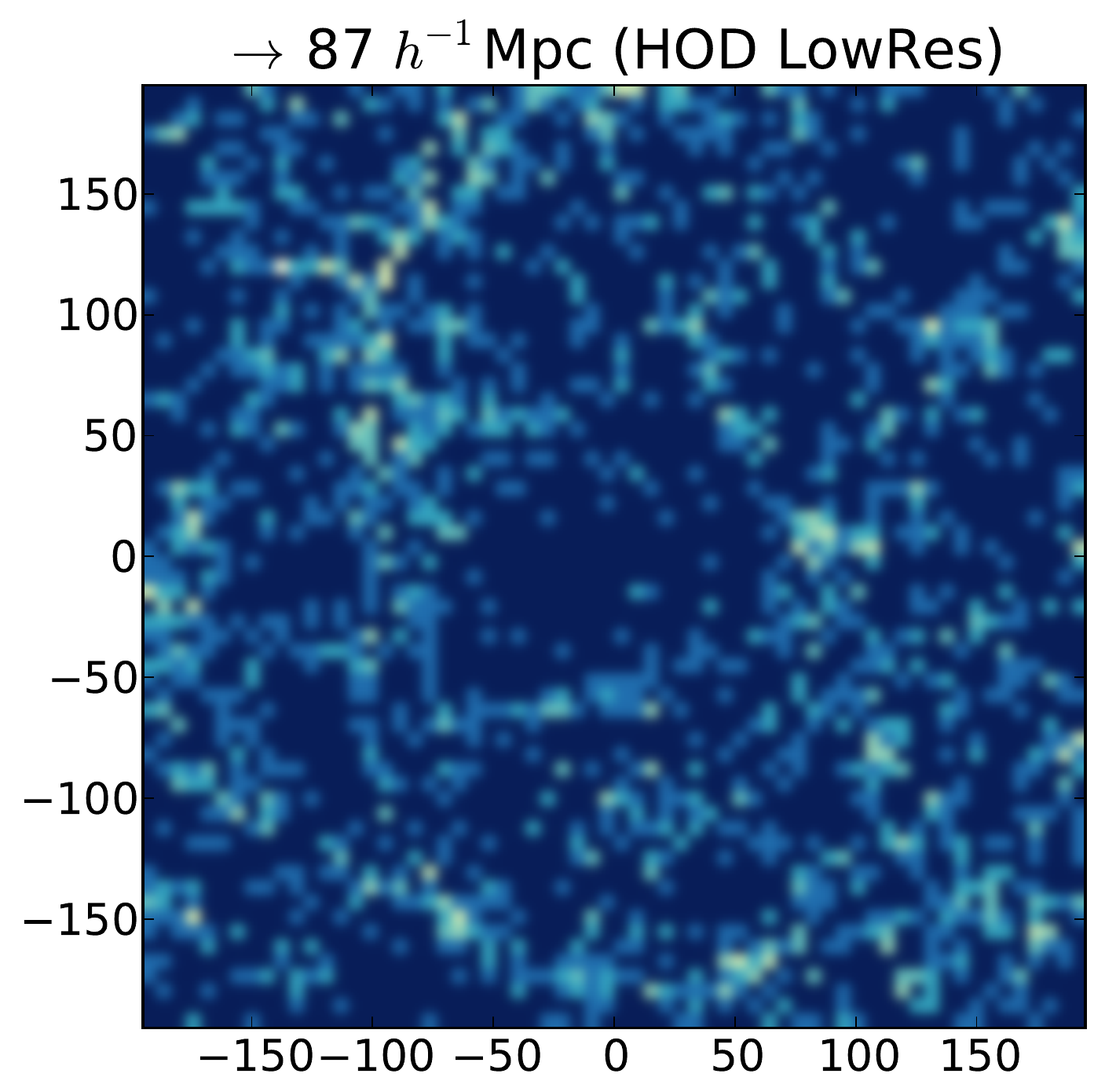}&
\includegraphics[width=0.7\columnwidth, angle=0]{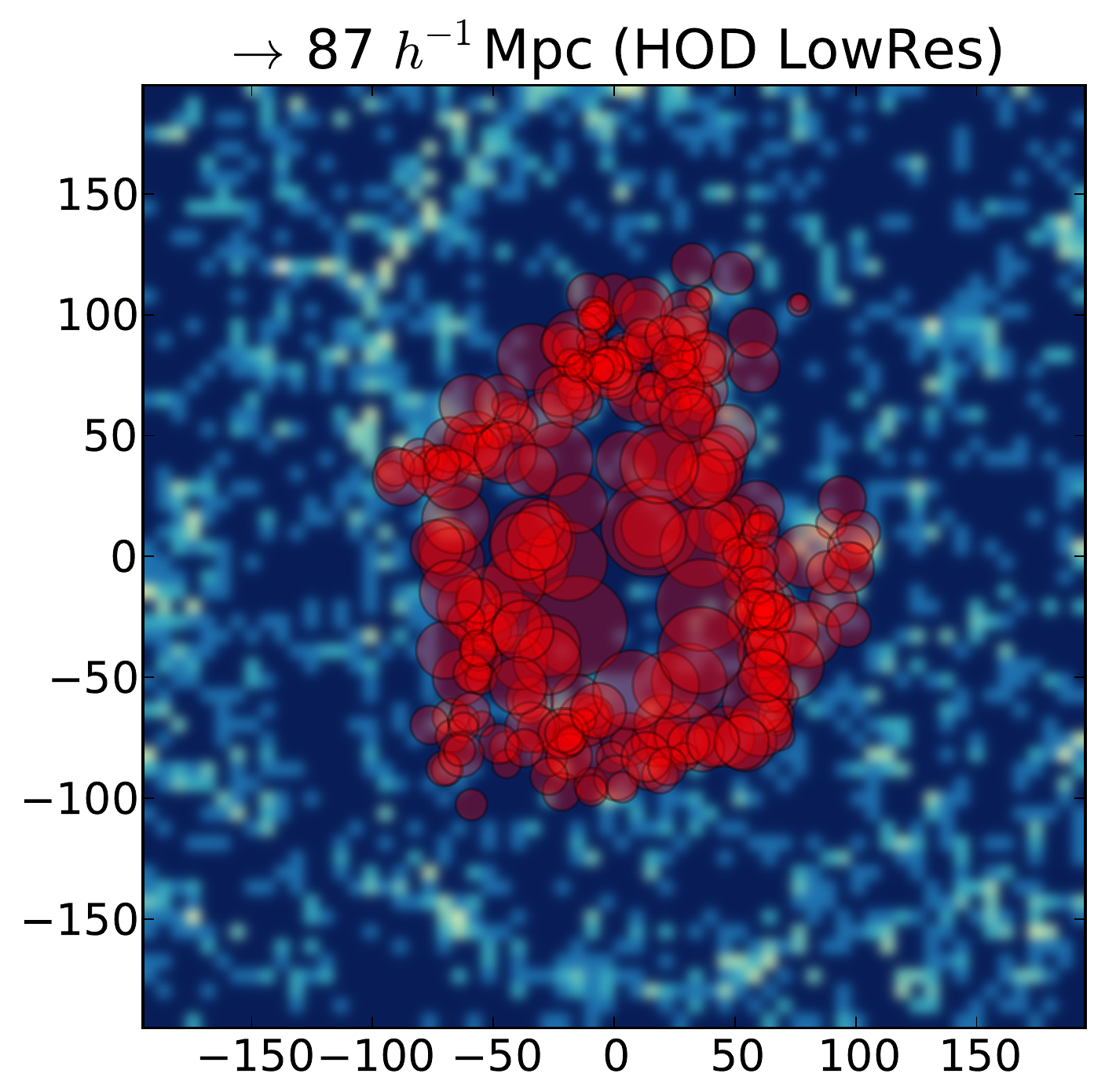}&
    \includegraphics[width=0.7\columnwidth, angle=0]{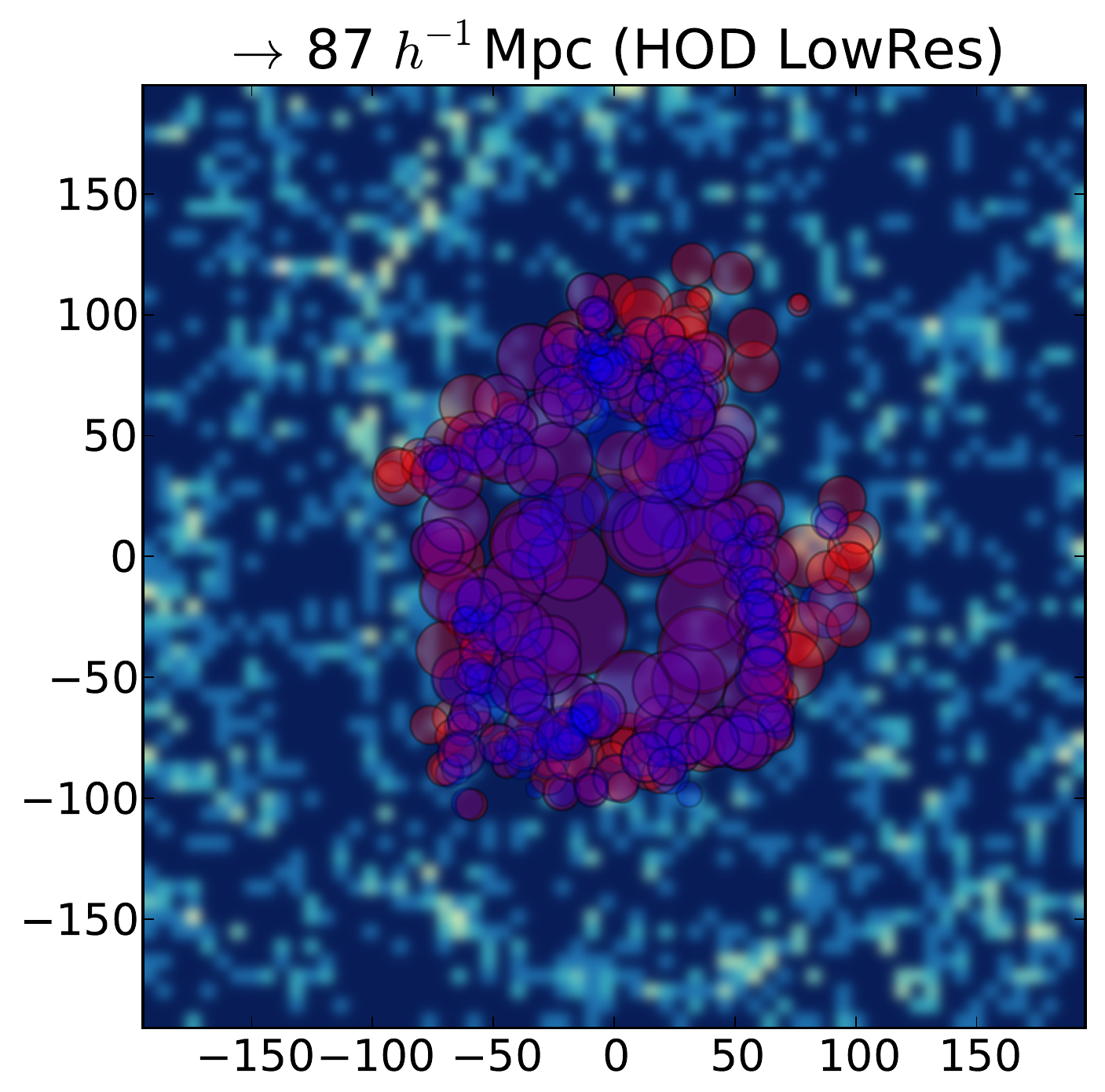}\\
  \end{tabular}
 \caption{
 Visual impression of the matching procedure for void catalogues. Left frame shows a slice of the density field without peculiar velocities. Central frame shows the same slice but with peculiar velocities and the void found in the slice (red). We represent the Voronoi cells constituting the void as small spheres, each with area related to the volume of the cell and center defined by the macrocenter of the cell. A visual comparison of the density field between the two panels illustrates the effect of peculiar velocities: as expected, structures are slightly enhanced. Right panel shows the void found in the density field without peculiar velocities (blue). The representation of voids is on the $x-y$ plane, units are \hmpc, and the slice is along the line of sight. Top row represents a void from the \textit{HighRes} sample, bottom row a void from the \textit{LowRes} sample.}
   \label{fig: void example}
\end{figure*}

\renewcommand{\tabcolsep}{0cm}
\begin{figure*}
\begin{tabular}{ccc}
  \includegraphics[width=0.7\columnwidth, angle=0]{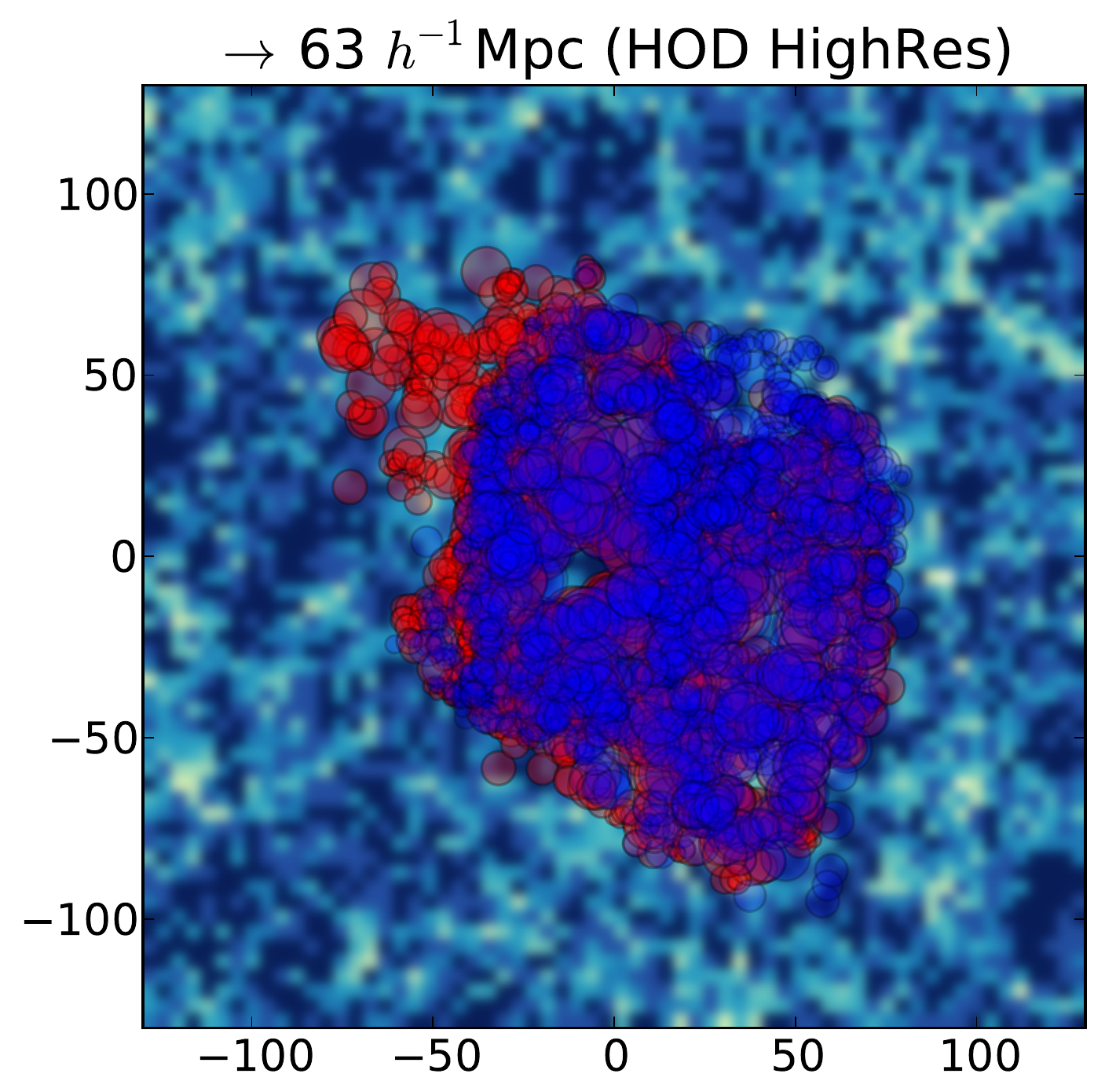}&
\includegraphics[width=0.7\columnwidth, angle=0]{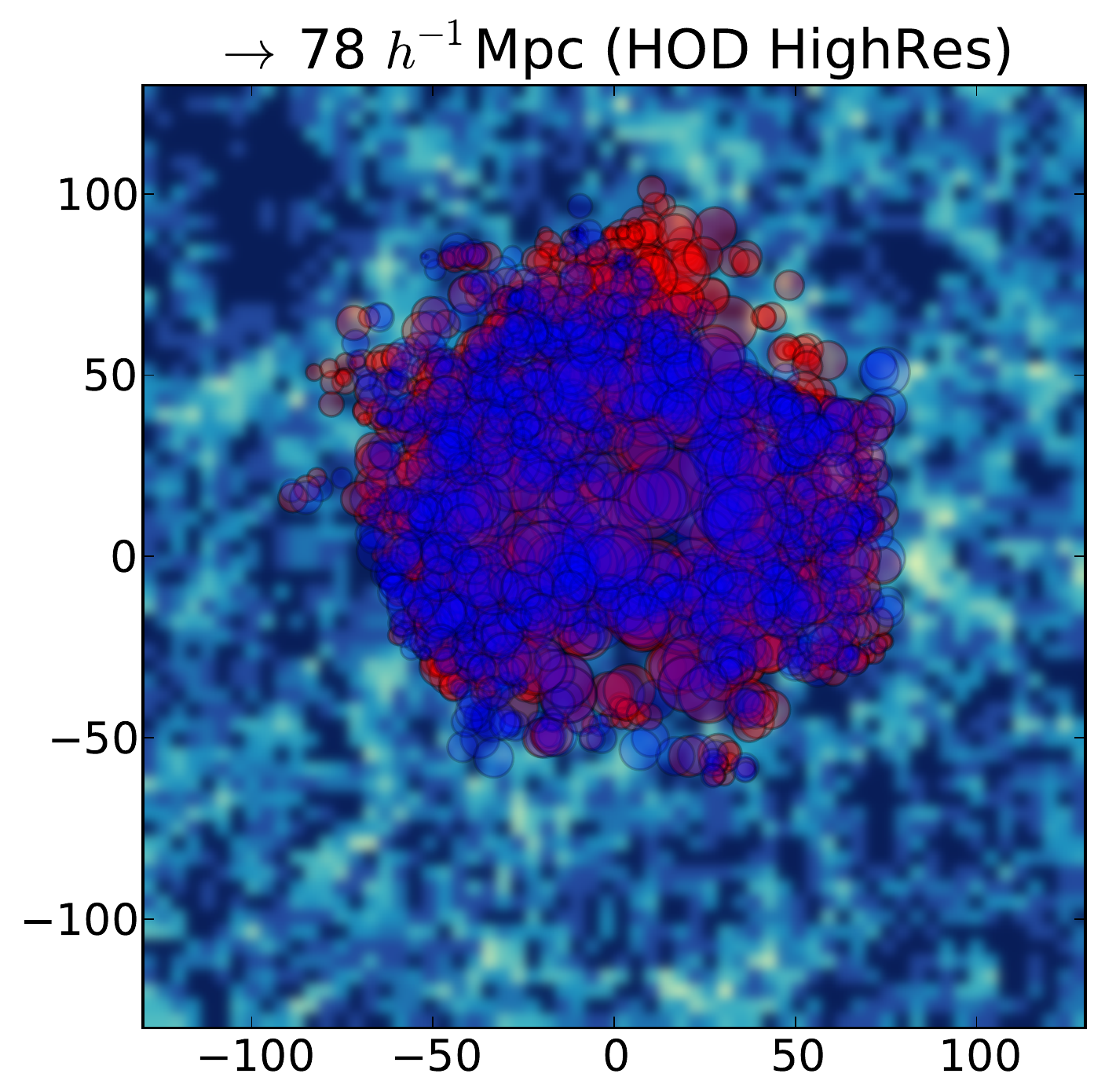}&
    \includegraphics[width=0.7\columnwidth, angle=0]{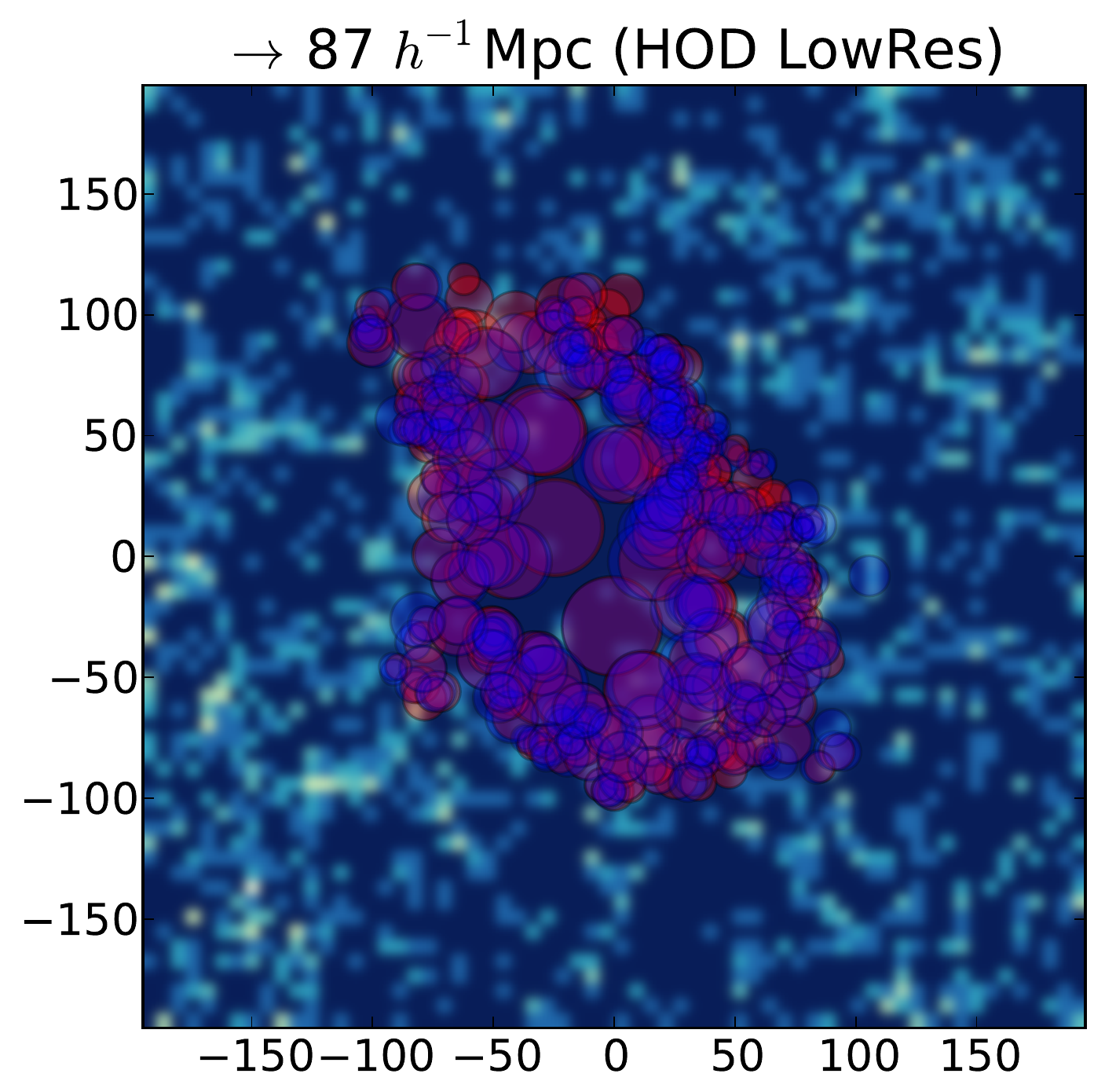}\\

  \end{tabular}
   \caption{
Slice of the density field in the $y-z$ plane, the line of sight is thus towards the left, qualitatively showing the effect of peculiar velocities on voids; the slice is with peculiar velocities. According to the color code in Figure \ref{fig: void example}, we show the void found in the density field with peculiar velocities in red, and the void found in the density field without peculiar velocities in blue. Left and central panel represent voids from the \textit{HighRes} sample, right represents a void from the \textit{LowRes} sample. We see that for these particular voids peculiar velocities only affect the detailed outer structure of the void, with minor effect on global properties such as radius and position.}
   \label{fig: peculiar velocity}
\end{figure*}

\subsection{The matching algorithm}
To analyse the effects of peculiar velocities on the void detection, we use the matching algorithm provided with the \vide toolkit, able to compare two voids catalogues in the most efficient way for our purposes. 

A simple but crucial point for our analysis is the choice of \textit{the catalogue with peculiar velocities as a base catalogue} for the comparison. This choice is instrumental, since it considers the correct perspective for the void finding: when we find voids in real surveys we observe them \textit{with }peculiar velocities.
Thus any study of peculiar velocities has to use the information we have as a starting point, so that the results of the analysis can be applied to a real galaxy survey, where we only have voids with peculiar velocities. 

For each void in the peculiar velocity catalogue, the matching algorithm selects all possible match candidates with centres lying in the Voronoi volume of the void. The matching method uses the unique particle IDs to find matches; it takes as the best match the candidate void with the highest number of shared particles. 

We use the matching methodology to perform two related analyses: first we apply the matching algorithm to check which voids found in the peculiar velocity mock correspond to voids in the mock without velocities (see Section 3). The unmatched voids are highly affected by peculiar velocities, at the point that they do not clearly correspond to any real-space void counterpart\footnote{With the void definition we are using. Of course, with a different definition of voids, such as the one proposed by \cite{Lavaux2010}, that is based on a Lagrangian orbit reconstruction, the situation might change.}and may be created by the effects of peculiar velocities. 

With the aim of characterizing the unmatched voids, we analyse their properties (such as ellipticity, density contrast or radius). Finally we disregard voids that are in the catalogue without peculiar velocities but not found in the catalogue with peculiar velocities, because in any way we will never be able to detect them --- the fact that we do not find them in the peculiar velocity mock means that velocities have erased them. Then we follow this analysis with a second study based on the relative properties of the matched voids (see Section 4), to measure the impact of velocities on a one-to-one basis. To assess the impact of velocities we thus consider the unmatched voids, as well as the properties of the matched voids, that is voids we measure in redshift space corresponding to voids without peculiar velocities.

The considered approaches are targeted to provide guidelines on the effect of peculiar velocities to all applications using voids from real data. 
As discussed above, for such applications, to avoid Poisson noise effects, voids with radii below the mean particle separation are already excluded, namely lower than 8 \hmpc for the \textit{HighRes} sample and 15 \hmpc for the \textit{LowRes} density sample.

\subsection{Example comparison}
To obtain a high quality one-to-one void comparison we use the matching routine of \vide described above. Nevertheless, it is very beneficial to obtain a good visualisation of voids, to check the behaviour of the matching algorithm and to understand the behaviour of voids when peculiar velocities are added. 

Since peculiar velocities will impact the detailed shape of voids, we represent them visually using the Voronoi cells constituting the void. Each cell is represented as a sphere, its area is related to the volume of the cell, and the center of the sphere is the particle position. The algorithm able to represent voids following this idea is part of the public void finder \vide\citep{Sutter2014a}\footnote{At \url{ http://bitbucket.org/cosmicvoids/vide_public}.}.

Although the Voronoi cells are not spherical, this constitutes an approximation that allows to observe the shape of voids in a particularly effective way. Figure \ref{fig: void example} shows the ability of the matching algorithm: it gives a visual impression of the quality of the matching on individual voids, allowing to see the effect of peculiar velocities on a one-to-one basis. The representation shows the amount of shape variation of voids, thus serving as a guide for the analysis of the peculiar velocities effect: we observe the slice and the matching voids from the two different catalogues. 

To illustrate the effect of velocities along the line of sight, we show in Figure \ref{fig: peculiar velocity} examples for the effect of velocities for a few voids. In this case, the line of sight is towards the left, \textit{i.~e.} we represent the voids on the $y-z$ plane. In Figures \ref{fig: void example} and \ref{fig: peculiar velocity} we qualitatively observe minor perturbations on the detailed void perimeter, for these particular example voids the effect of peculiar velocities on global properties such as radius and position is negligible.

\section{Results: void survival rate}

As discussed in the previous section, we used the catalogue with peculiar velocities as a base catalogue for the comparison. In order to extract cosmological information, we would like to know which voids are the most affected by peculiar velocities. In such voids, the effect of cosmology would be  dominated by the effect of peculiar velocities. If we found a way to characterize peculiar-velocity dominated voids, we could wisely exclude them from the analysis, as the signal-to-noise ratio for cosmological information with these voids would be low. In the next Subsection, we use the matched fraction of voids and the ellipticities to characterise the effect of peculiar velocities on cosmic voids.  

\subsection{Matched fraction\label{sec: matched fraction}}

A void that is weakly affected by peculiar velocities will have similar properties in both simulations --- a direct way to exclude the most affected voids is to consider the matched and unmatched voids. 

\begin{figure}
\begin{tabular}{c}
  \includegraphics[width=1\columnwidth, angle=0]{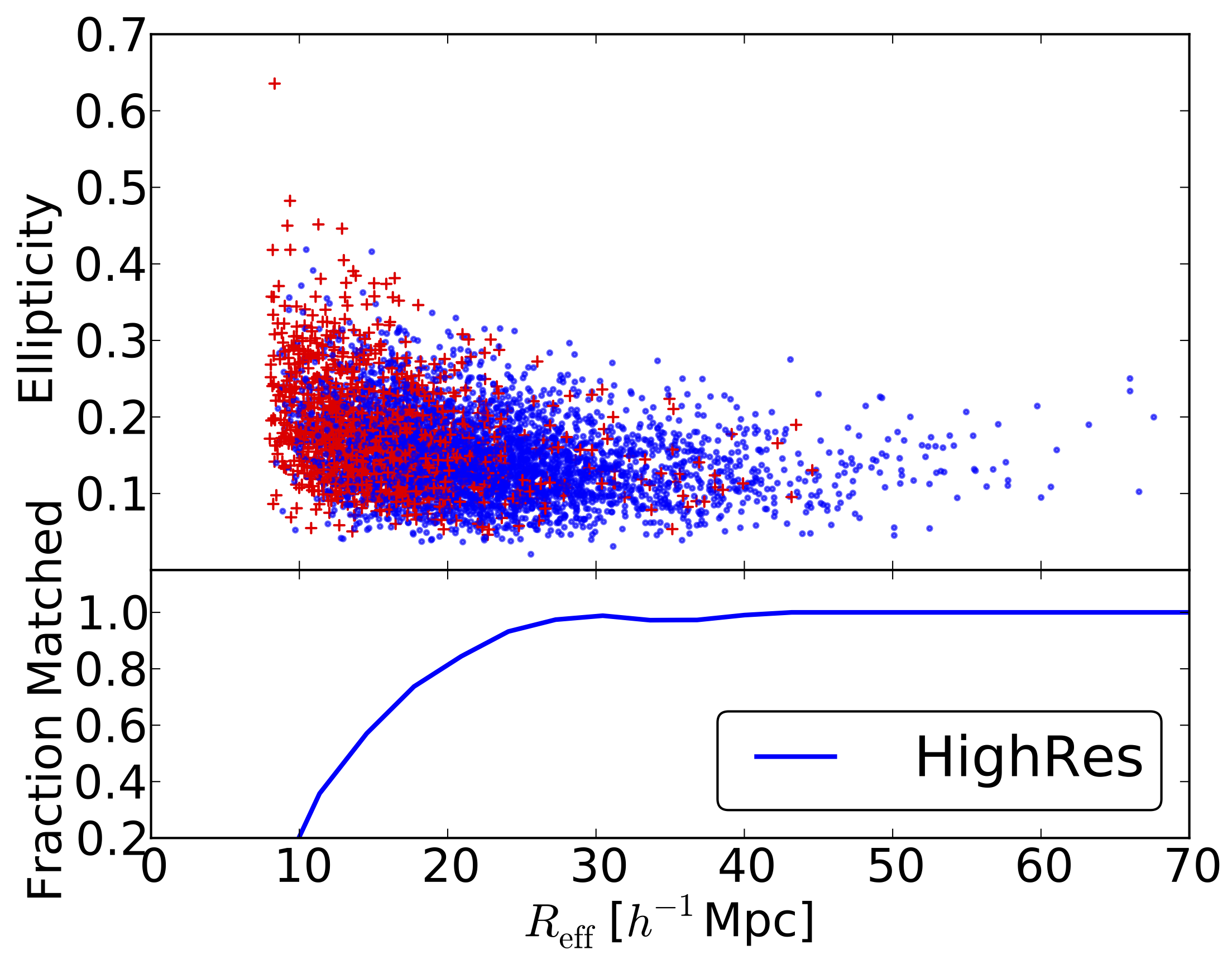}\\
    \includegraphics[width=1\columnwidth, angle=0]{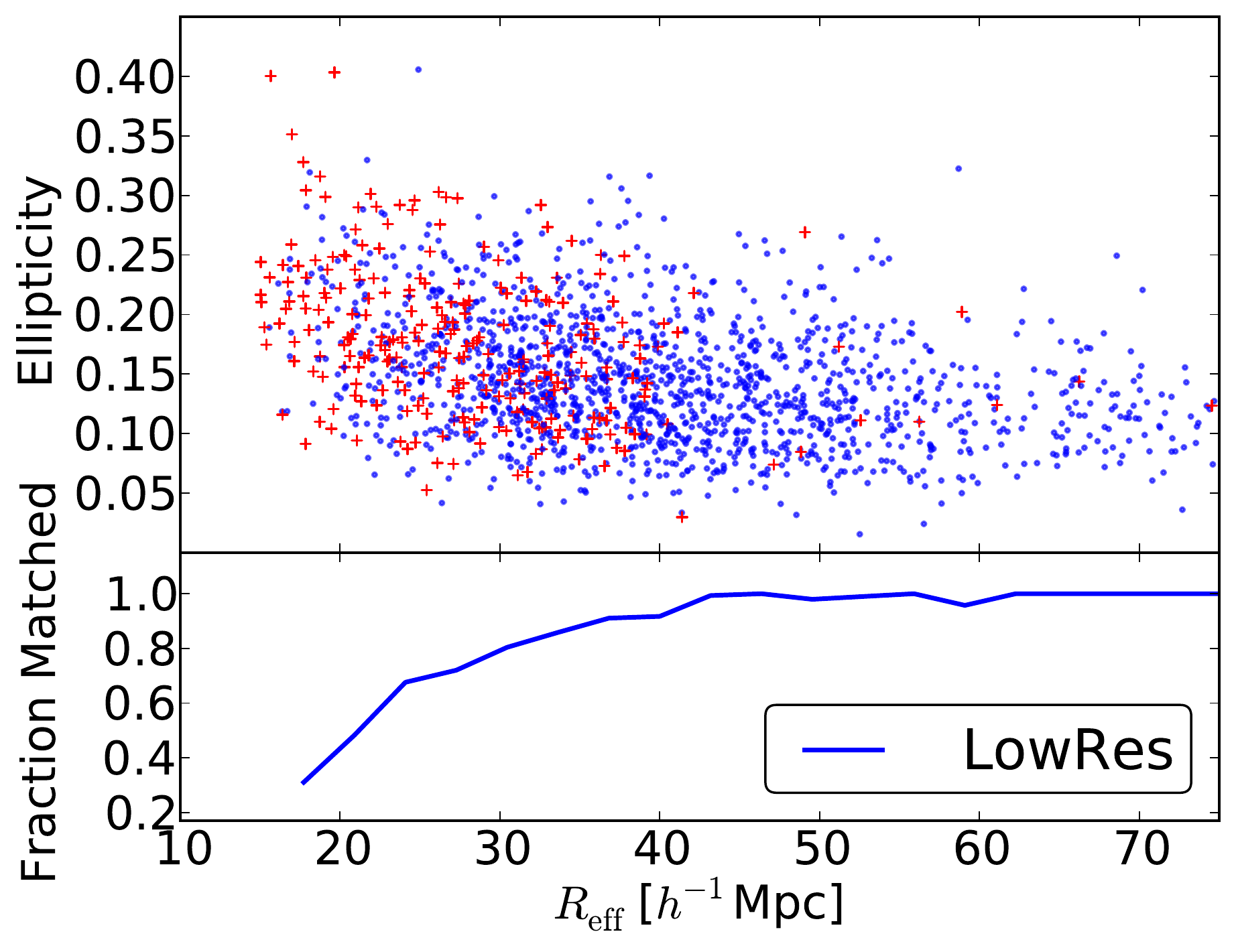}\\  
  \end{tabular}
 \caption{We represent in the top plot of each panel the matched and unmatched voids from the catalogues in the radius-ellipticity plane for both the \textit{HighRes} and \textit{LowRes} sample. Red crosses represent unmatched voids; blue dots represent matched voids. The bottom plot of each panel shows the fraction of matched voids.
Smaller voids can be significantly contaminated by peculiar velocities --- they include a large fraction of the unmatched voids and are peculiar velocity-dominated (while here is no clear distinction in ellipticity). We show that voids smaller than $\sim$ 20 \hmpc for the \textit{HighRes} sample and smaller than $\sim$ 35 \hmpc in the \textit{LowRes} sample are more affected by peculiar velocities. These limits in radius roughly correspond to twice the mean particle separation. However, we also note that there are a population of well-matched, minimally-affected voids at all scales.}
   \label{fig: void unmatched}
\end{figure}

Taking the peculiar velocity catalogue as a basis for the comparison, unmatched voids from the non-peculiar sample are the most affected by peculiar velocities. For this analysis we considered ellipticity and radius of voids, to understand which properties of the unmatched voids are a feature of a velocity-dominated void. Figure \ref{fig: void unmatched} shows the matched and unmatched voids from the catalogues with and without peculiar velocities in the radius-ellipticity plane for both the \textit{HighRes}  and \textit{LowRes} sample. The matching is worse for voids with radii lower than $\sim$ 20 \hmpc for the \textit{HighRes} sample and lower than $\sim$ 35 \hmpc for the \textit{LowRes} sample, indicating that, when finding small voids, results can be strongly affected by peculiar velocities. This seems particularly logical: we might have expected, a priori, that small voids are the most affected by changes in shape due to the velocities, which affects the way \vide defines the Voronoi cells and selects them as belonging to a void. Additionally, since small voids are more likely to be found in higher-density environments, they have higher ridges \citep{Hamaus2014}. Thus they are more affected by the peculiar velocities of the high-density structures forming the walls.
The finding of these voids is affected by peculiar velocities, thus their features are peculiar-velocity dominated. Figure \ref{fig: void unmatched} shows that, while some properties of voids (such as ellipticity) are affected by peculiar velocities, the effect is not the same for all voids, and does not show any preferential feature that can be used to identify peculiar velocity-dominated voids. On the contrary, the radii of voids seem to allow to identify unmatched voids among the smallest ones: properties of smaller voids are more likely to be dominated by peculiar velocities.

When extracting cosmological information from the unmatched voids, the relative strength of peculiar velocities can lead to biases or systematics: the signal-to-noise is low due to the strong impact of velocities. With the aim of extracting cosmological information from voids --- for instance using Alcock-Paczy\'nski test --- this consideration should be taken into account, as it would greatly improve the result: low radii voids should be wisely excluded from the analysis to maximize the signal-to-noise.

The bottom plot of each panel in Figure \ref{fig: void unmatched} shows the fraction of matched voids, confirming the guidelines we obtain from the top plot: for small voids the matching between voids without and with peculiar is less reliable. Thus when we find voids in redshift space, the smallest voids are the ones most affected  by peculiar velocities and for which the matching is often worse, in that case the fraction of matched voids is lower. The fraction of matched voids is higher than 80\% for voids bigger than 20 \hmpc for the \textit{HighRes} sample and 30 \hmpc for the \textit{LowRes} sample. 

The use of cuts when working with voids can be a powerful tool to ensure a reduction of systematics errors. We point out that, depending on applications, the considered cut might vary. This does not apply to cuts aimed to avoid Poisson dominated voids (namely \textit{once} the mean particle separation) but to other following cuts. 

\subsection{Cuts and efficiency}

The previous Section showed that small voids are peculiar-velocity dominated objects: the use of a cut that excludes voids of small radius could reduce the effect of peculiar velocities on voids. For this purpose one would ideally need to find the trade-off between reducing intrinsic variance due to peculiar velocities and increasing statistical variance (due to the fact that we would remove voids). Nevertheless, the motivation for the cut might change, thus changing the cut itself. For example if we wanted to study the effect of peculiar velocities and eventually measure the peculiar velocity around voids, the voids with higher information would be the ones \textit{below} the radius cut, \textit{i.~e.~}small voids. 

From these considerations about unmatched voids, if the detailed structure of a void is necessary for the analysis, one should avoid using the smallest voids, to be more precise voids smaller than \textit{twice the mean particle separation}, since they may be contaminated by peculiar velocities.

For the \textit{HighRes} and \textit{LowRes} samples, this would mean excluding voids smaller than 16 \hmpc and 30 \hmpc respectively. While this cut might seem drastic for current surveys (such as SDSS DR7) and might excessively reduce the number of voids, it can be a good prescription to be adopted for future applications of the \ap.

\begin{figure}
\begin{tabular}{c}
  \includegraphics[width=1\columnwidth, angle=0]{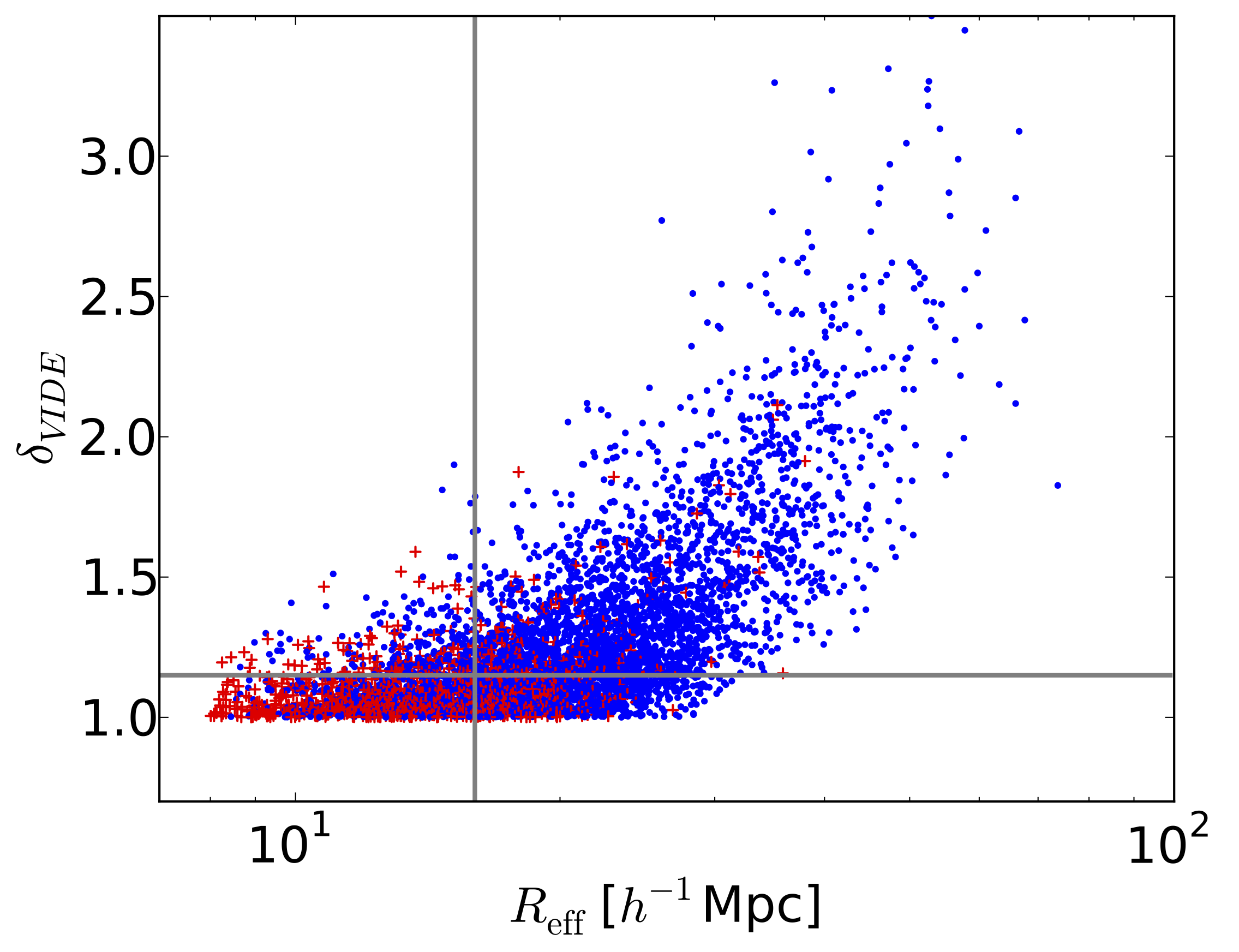}\\
    \includegraphics[width=1\columnwidth, angle=0]{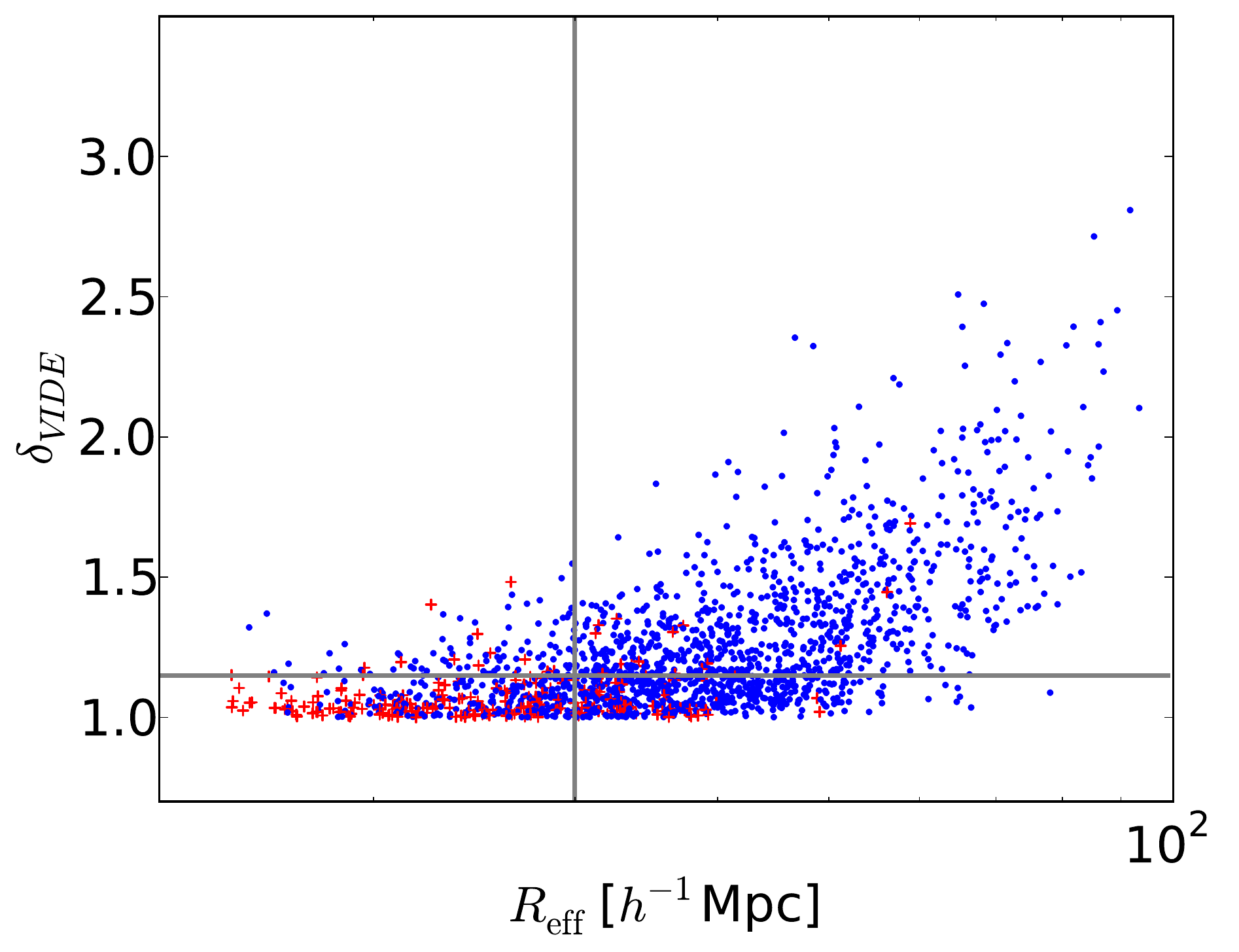}\\  
  \end{tabular}
 \caption{We show with blue dots the voids matched between the peculiar velocity and no-peculiar velocity catalogues, and with red crosses the unmatched. In both the \textit{HighRes} (left) and \textit{LowRes} (right) cases we show with grey lines an example cut based on the radius and density contrast that removes a large fraction of unmatched voids (see Tables \ref{tab: cut high} and \ref{tab: cut low}). The considered cuts only use information of the catalogue with peculiar velocities and are therefore applicable to data from real surveys.}
   \label{fig: Density Contrast}
\end{figure}

As Figure \ref{fig: Density Contrast} shows, it is possible to further isolate unmatched voids by applying, \textit{additionally to the cut in radius}, a cut based on the density contrast of voids (using the density contrast as defined by \zobov and thus used by \vide, which is given by the ratio of the minimum density along the ridge of the void versus the minimum density in the void). We see that unmatched voids (\textit{i.~e.~}, those strongly affected by peculiar velocities) cluster in the small radii, low density contrast corner of the diagrams --- which seems intuitively consistent. Voids with low density contrast are shallower, thus the effect of peculiar velocities on such voids is stronger.

As an example, we show in Table \ref{tab: cut high} and Table \ref{tab: cut low} the results of the example cuts shown in Figure \ref{fig: Density Contrast}: excluding voids with radii below twice the mean particle separation in radius and with density contrast below $1.15$. With these cuts the remaining unmatched voids are $\sim$5\% for the high density sample and $\sim$2\% in the low density sample. We point out that the unmatched voids are legitimate voids, the only reason we consider their exclusion is because they are more affected by velocities.

\begin{table}[h]
\centering 
\begin{tabular}{@{\extracolsep{\fill}} c c c c c }
\caption{Example of removal process with cuts for the high density sample.\label{tab: cut high}}
\\
\hline    
Cut& \%Match \hspace{4pt} & \%Unmatch \hspace{4pt} & \%Total \hspace{4pt} & \%Remaining  \\   
&  Removal& Removal &  Removed &  Unmatch \\   
\hline                       
2$R_{\rm{mps}} $& 20.0 &  63.9& 28.6 &  9.98\\
1.15$\delta_{VIDE}$&  39.0 & 77.4 & 46.6  & 8.34\\
Both&  44.6& 87.5&53.1& 5.26\\ 
\hline    
 \end{tabular}
 \end{table}

 \begin{table}[h]
\centering 
\begin{tabular}{@{\extracolsep{\fill}} c c c c c }
\caption{Example of removal process with cuts for the low density sample.\label{tab: cut low}}
\\
\hline    
Cut& \%Match \hspace{4pt} & \%Unmatch \hspace{4pt} & \%Total \hspace{4pt} & \%Remaining  \\   
&  Removal& Removal &  Removed &  Unmatch \\   
\hline                       
2$R_{\rm{mps}} $& 19.6 &  67.0& 25.4 &  5.39\\
1.15$\delta_{VIDE}$&  43.3&86.3  & 48.5& 3.24\\
Both&  49.1& 92.5&54.4& 2.02\\ 
\hline    
 \end{tabular}
 \end{table}

Nevertheless, for current datasets, these cuts also remove many  unaffected voids. In Table \ref{tab: cut high} and \ref{tab: cut low} we see that the example cut leaves as wished a very small number of peculiar-velocity dominated voids (remaining unmatched), but at the expense of generally reducing the number of voids and in particular excluding many of the good voids (the fraction of matched removal is high in both cases).

To avoid this effect, we might want to consider cuts that avoid the exclusion of too many matched voids. A good trade-off between cutting too many voids and excluding as many non-matched voids as possible can be found using our matching algorithm: one can consider a cut minimising the number of removed matched voids and maximising the number of removed unmatched voids. We perform a minimisation taking into account these criteria in the parameter space of the density contrast and radius. 
More precisely, we define the \textit{efficiency} ($\eta$) of the cut as the ratio of the percentage of unmatched voids removed $N_{removed}^{unmatch}$ to the percentage of the matched voids kept $N_{kept}^{match}$: 
\begin{equation}
\eta=\frac{N_{removed}^{unmatch}}{N_{kept}^{match}} \label{eq: efficiency}
\end{equation}
Ideally, we would want the efficiency to be 1 (100\% of unmatched voids removed and 100\% of matched voids kept), thus, to find the best parameters for the cuts in radius and density contrast, we minimize the function:
\begin{equation}
f(\delta_{cut},R_{cut})=|1-\eta| \label{eq: function}
\end{equation}

\begin{figure}
\begin{center}
\begin{tabular}{c}

 \includegraphics[width=0.9\columnwidth, angle=0]{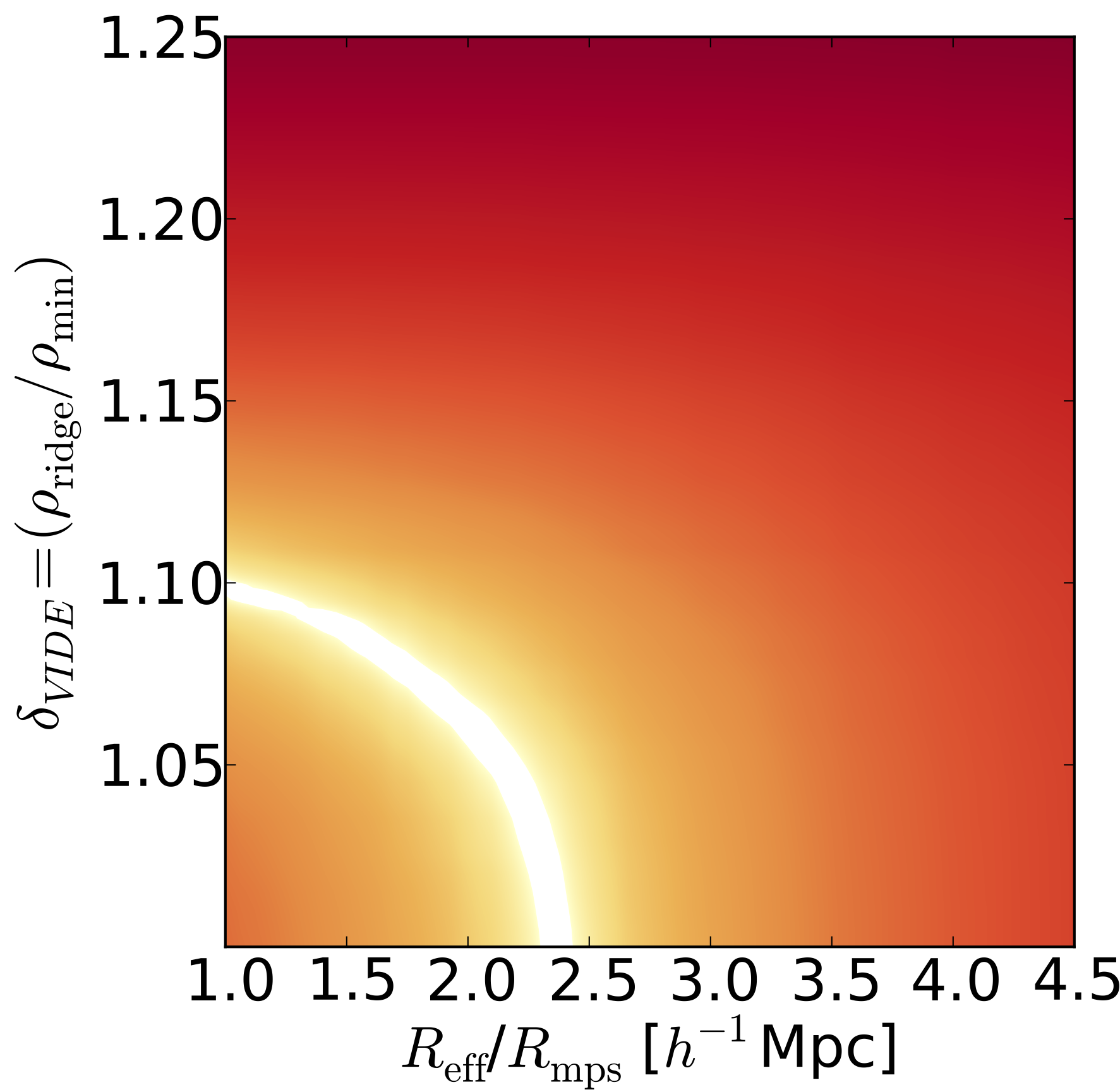}\\
   \includegraphics[width=0.9\columnwidth, angle=0]{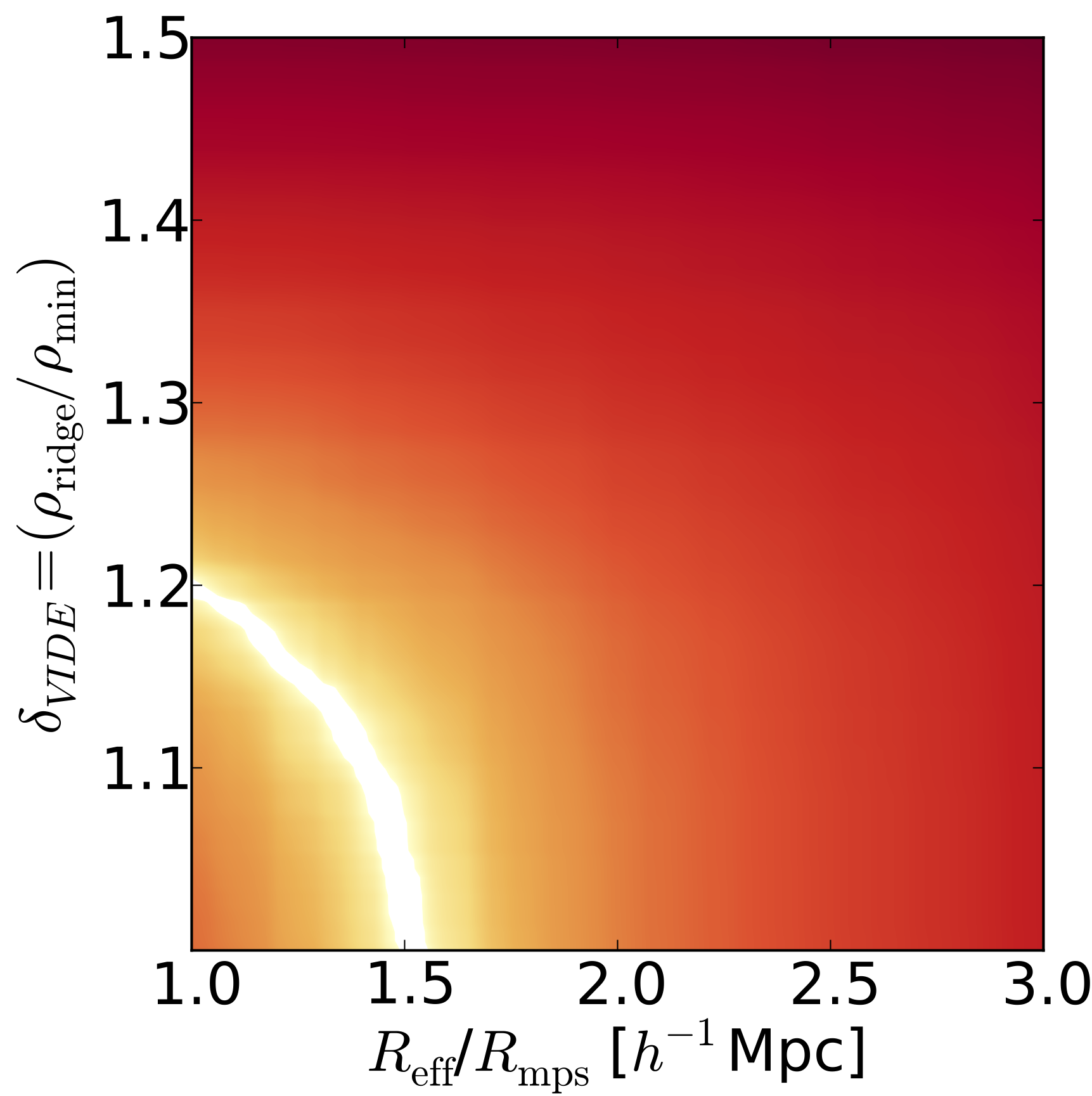}\\
  \end{tabular}
 \caption{Maximisation of the efficiency (see equations \ref{eq: efficiency} and \ref{eq: function}, we represent the function $f$) depending on the different values of cuts on density contrast and radius for the \textit{HighRes} sample (top) and the \textit{LowRes} sample (bottom). White shows the best efficiency zone for cuts. }
   \label{fig: parameter space}
   \end{center}
\end{figure}

\begin{figure}
\begin{center}
\begin{tabular}{c}
  \includegraphics[width=0.86\columnwidth, angle=0]{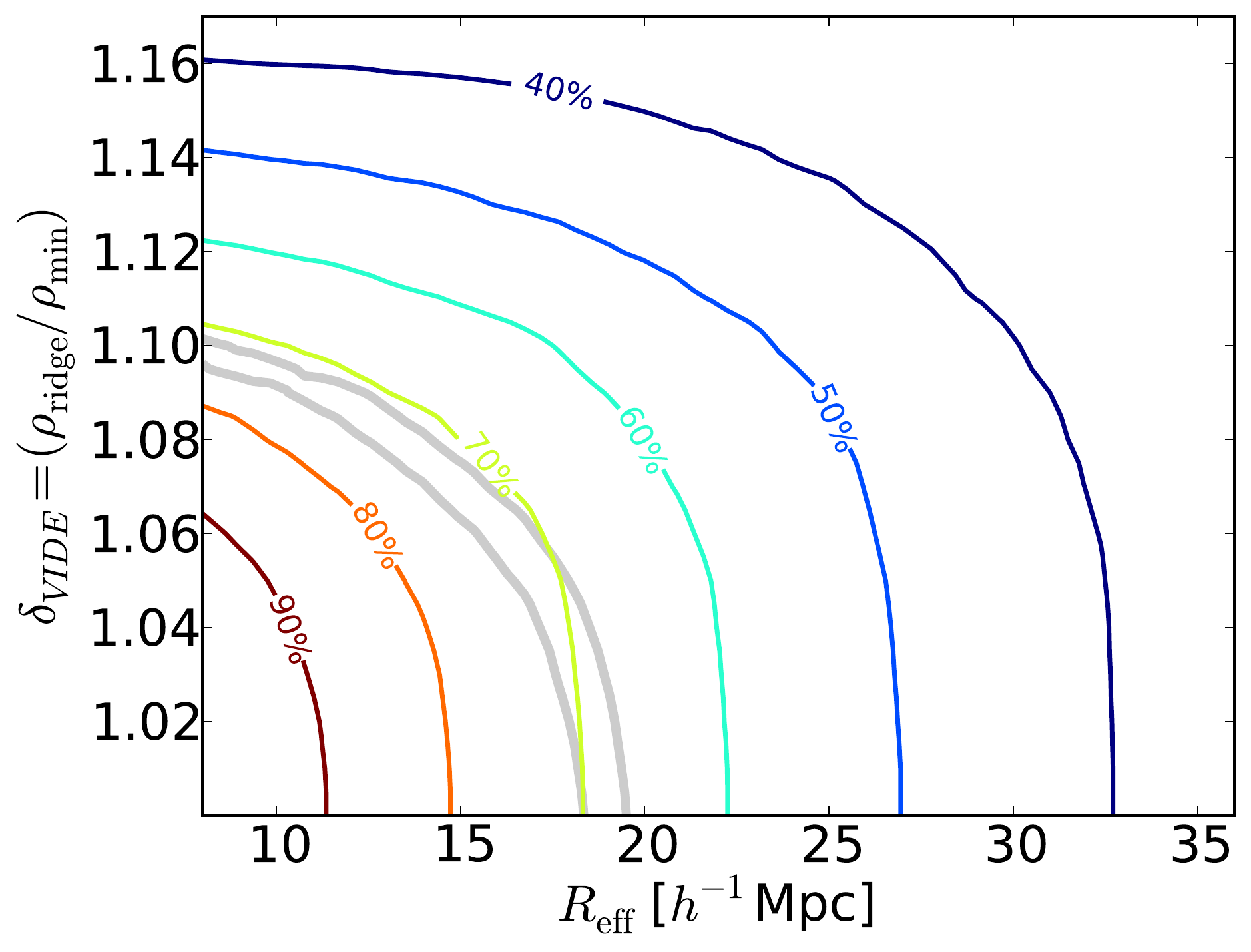}\\
     \includegraphics[width=0.86\columnwidth, angle=0]{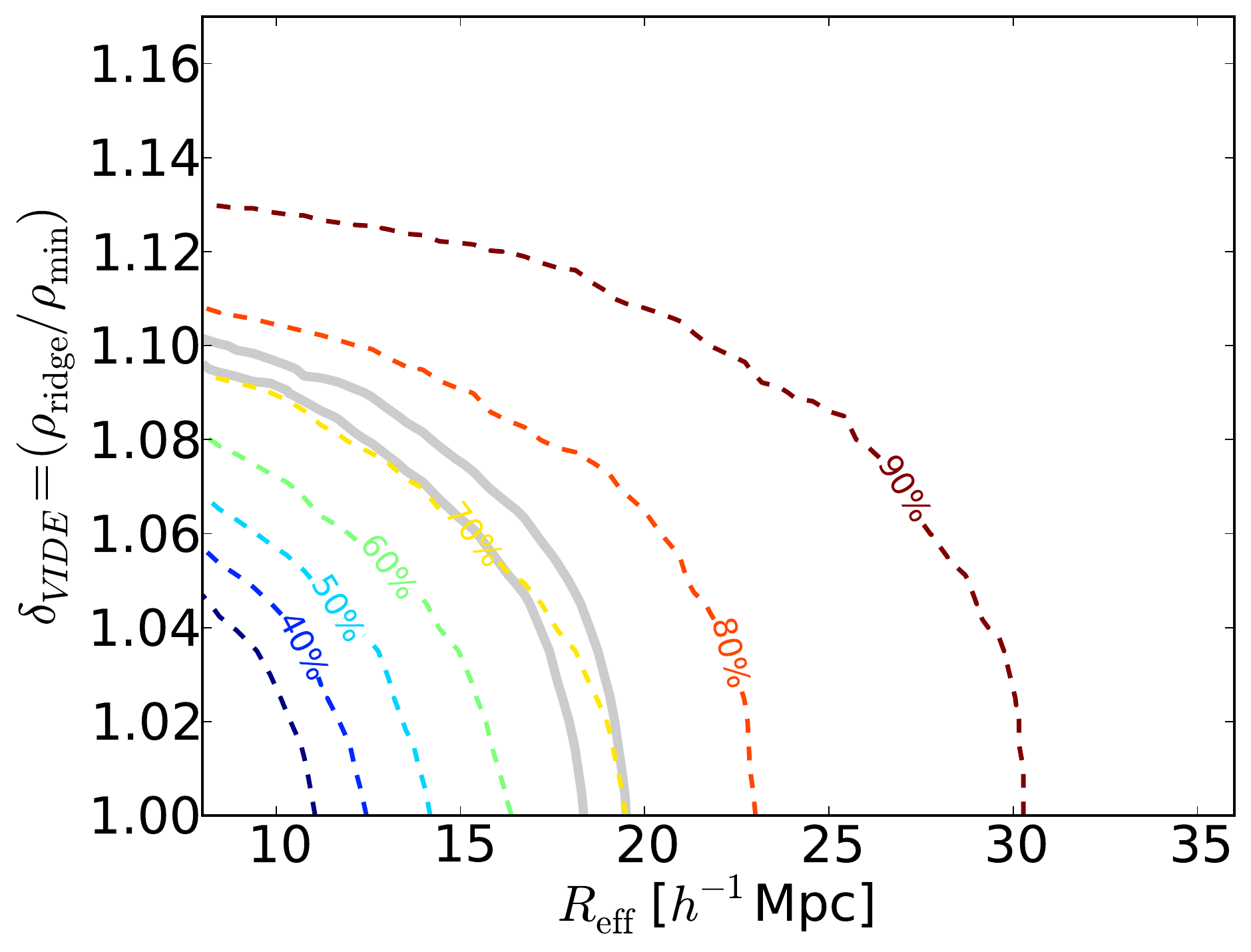}\\
  \end{tabular}
  \end{center}
 \caption{For the \textit{HighRes} sample: percentage of matched voids kept for each cut (top) 
 and percentage of voids dominated by peculiar velocities (unmatched voids) removed for each cut (bottom). For reference, grey thick lines show the ideal area for cuts (see Figure \ref{fig: parameter space}): a choice of parameters in this ideal zone typically gives $\sim 70\%$ of unmatched voids removed at the expense of $\sim 30\%$ of matched voids removed (corresponding to $\sim 70\%$ of matched voids kept).}
   \label{fig: percent high}
\end{figure}

\begin{figure}
\begin{center}

\begin{tabular}{c}
\includegraphics[width=0.86\columnwidth, angle=0]{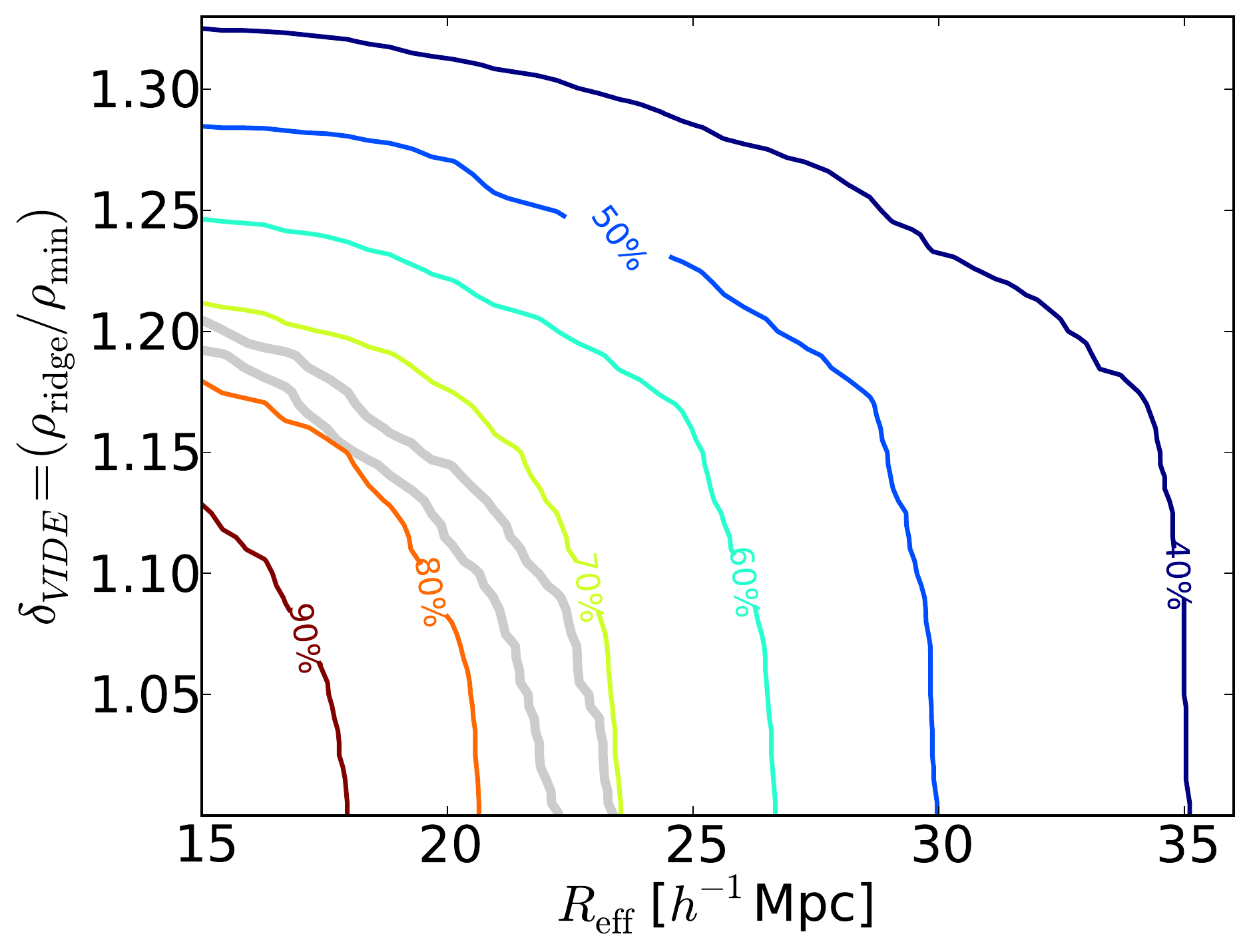}\\
 \includegraphics[width=0.86\columnwidth, angle=0]{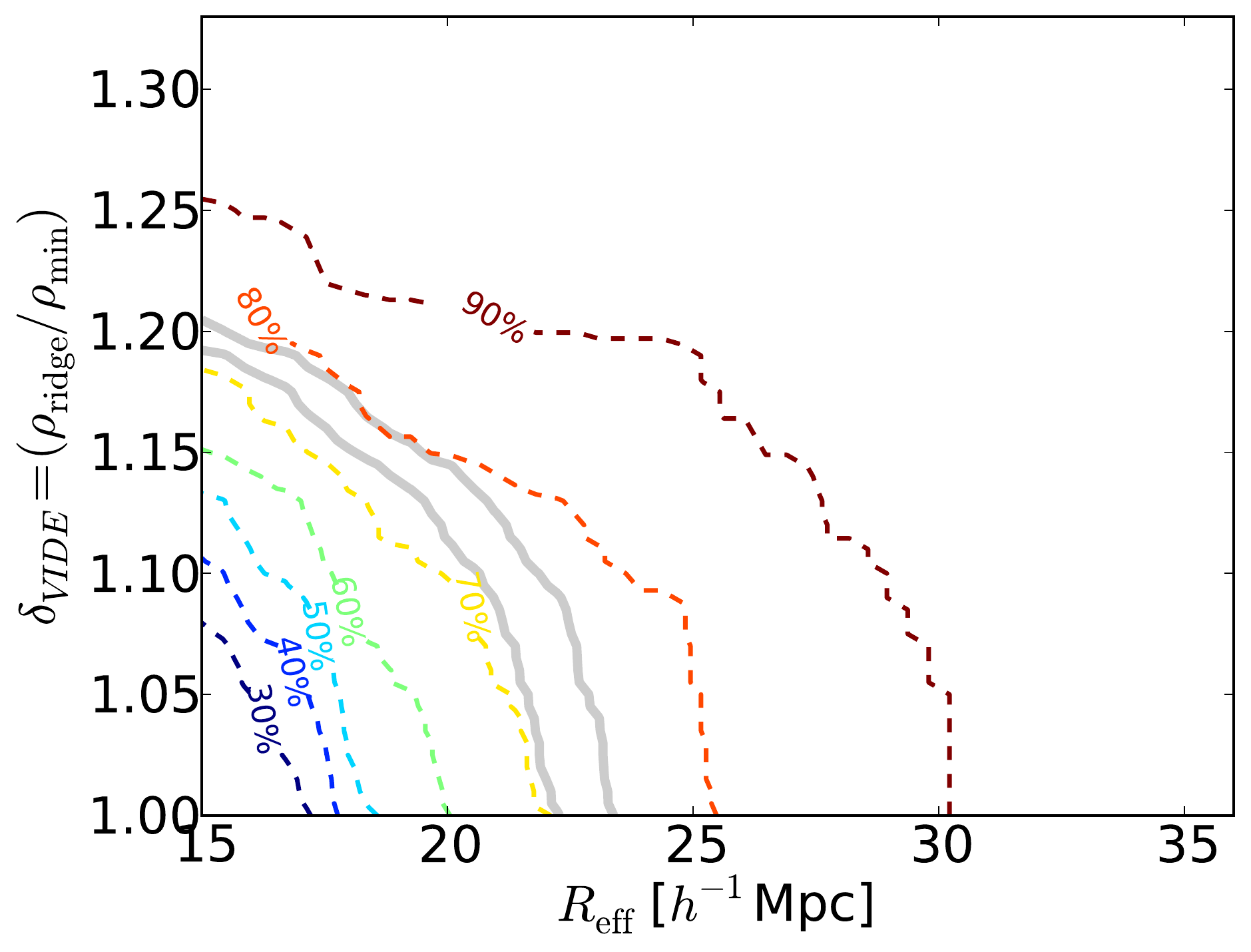}\\
  \end{tabular}
  \end{center}
 \caption{For the \textit{LowRes} sample: percentage of matched voids kept for each cut (top) 
 and percentage of voids dominated by peculiar velocities (unmatched voids) removed for each cut (bottom). For reference, grey thick lines show the ideal area for cuts (see Figure \ref{fig: parameter space}): a choice of parameters in this ideal zone typically gives $\sim 75\%$ of unmatched voids removed at the expense of $\sim 25\%$ of matched voids removed (corresponding to $\sim 75\%$ of matched voids kept).}
   \label{fig: percent low}
\end{figure}

\begin{figure}
\centering  
\begin{tabular}{c}
  \includegraphics[width=0.9 \columnwidth, angle=0]{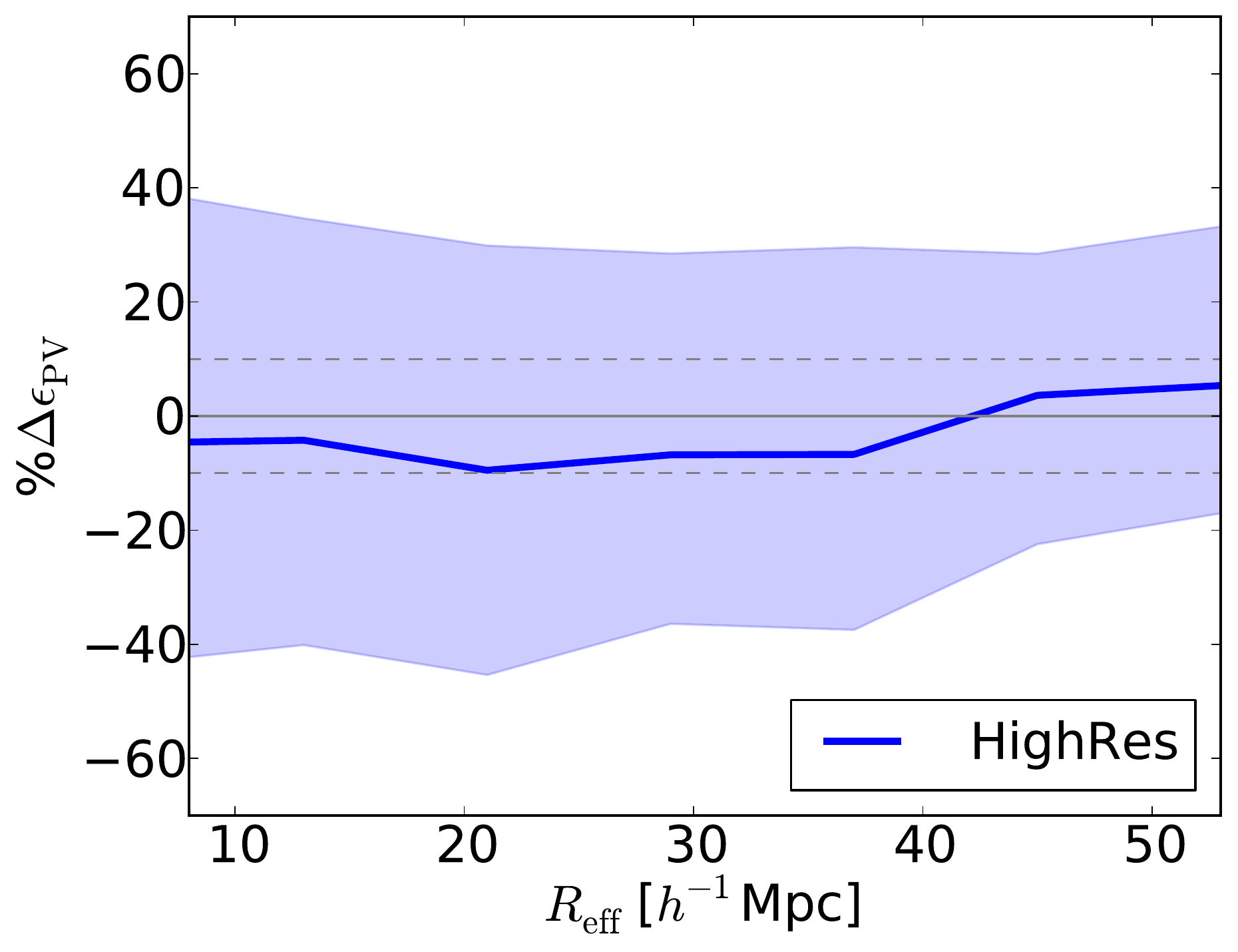}\\
  \includegraphics[width=0.9\columnwidth, angle=0]{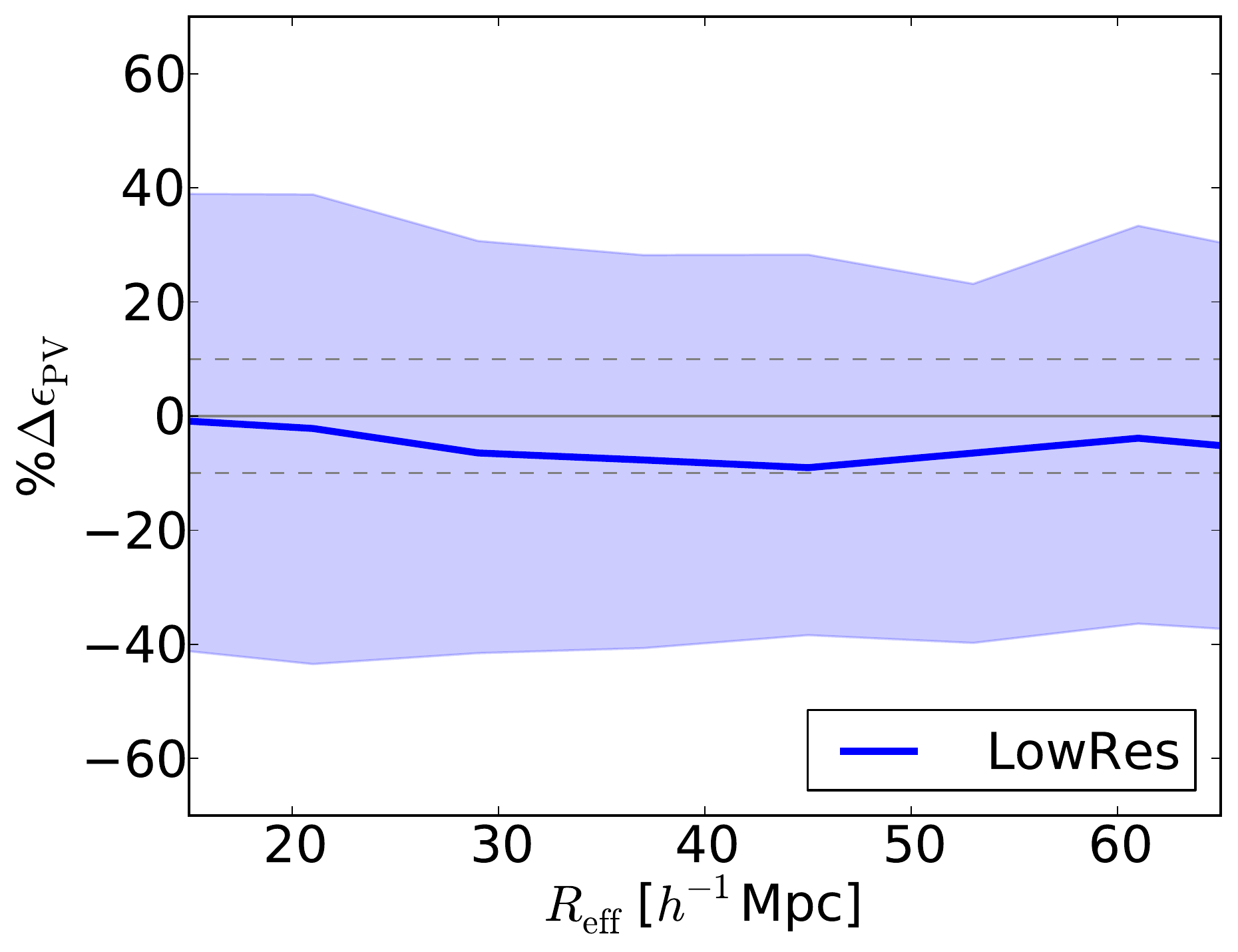}
  \end{tabular}
 \caption{Percentage ellipticity change due to peculiar velocities for the \textit{HighRes} sample (top) and for the \textit{LowRes} sample (bottom) for matched voids. Peculiar velocities contribute on average to an increase in the sphericity of voids. We find an average reduction of the ellipticity of voids at all scales (percentage ellipticity change lower than 10\% of the value, grey dashed line), except for the largest in the \textit{HighRes} sample, which are extended by a small percentage. The shaded regions are the $1\sigma$ Poisson uncertainties.}
 
   \label{fig: percentage}
\end{figure}

Figure \ref{fig: parameter space} shows the exploration of different values for the cuts for the high density sample (top) and for the low density sample (bottom).

We see an ideal zone in a curved region where the choice of parameters maximises the efficiency of cuts to master the effect of peculiar velocities on voids by excluding the most affected ones. In this zone the function $f$ is minimized, thus maximizing the efficiency. 

Additionally, we show in Figure \ref{fig: percent high} (\textit{HighRes} sample) and in Figure \ref{fig: percent low} (\textit{LowRes} sample) the percentage of matched voids kept for each cut (top) and the percentage of voids dominated by peculiar velocities that are removed for each cut (bottom). Using values from the ideal zone shown in Figure \ref{fig: parameter space} gives the best efficiency. A choice of parameters in this ideal zone typically gives for the \textit{HighRes} sample $\sim 70\%$ of unmatched voids removed at the expense of $\sim 30\%$ of matched voids removed; and for the \textit{LowRes} sample a removal rate of $\sim 75\%$ for the unmatched voids and $\sim 25\%$ for the matched. For reference, in case the choice of values was not in that region, Figures \ref{fig: percent high} and \ref{fig: percent low} show the percentage of voids after any cut on radius or density contrast. 

Current and future surveys may particularly profit of the application of such cuts, to obtain catalogues of voids with reduced systematics, particularly adapted for the extraction of the cosmological signal.

\begin{figure}
\begin{center}
\begin{tabular}{c}
  \includegraphics[width=0.86\columnwidth, angle=0]{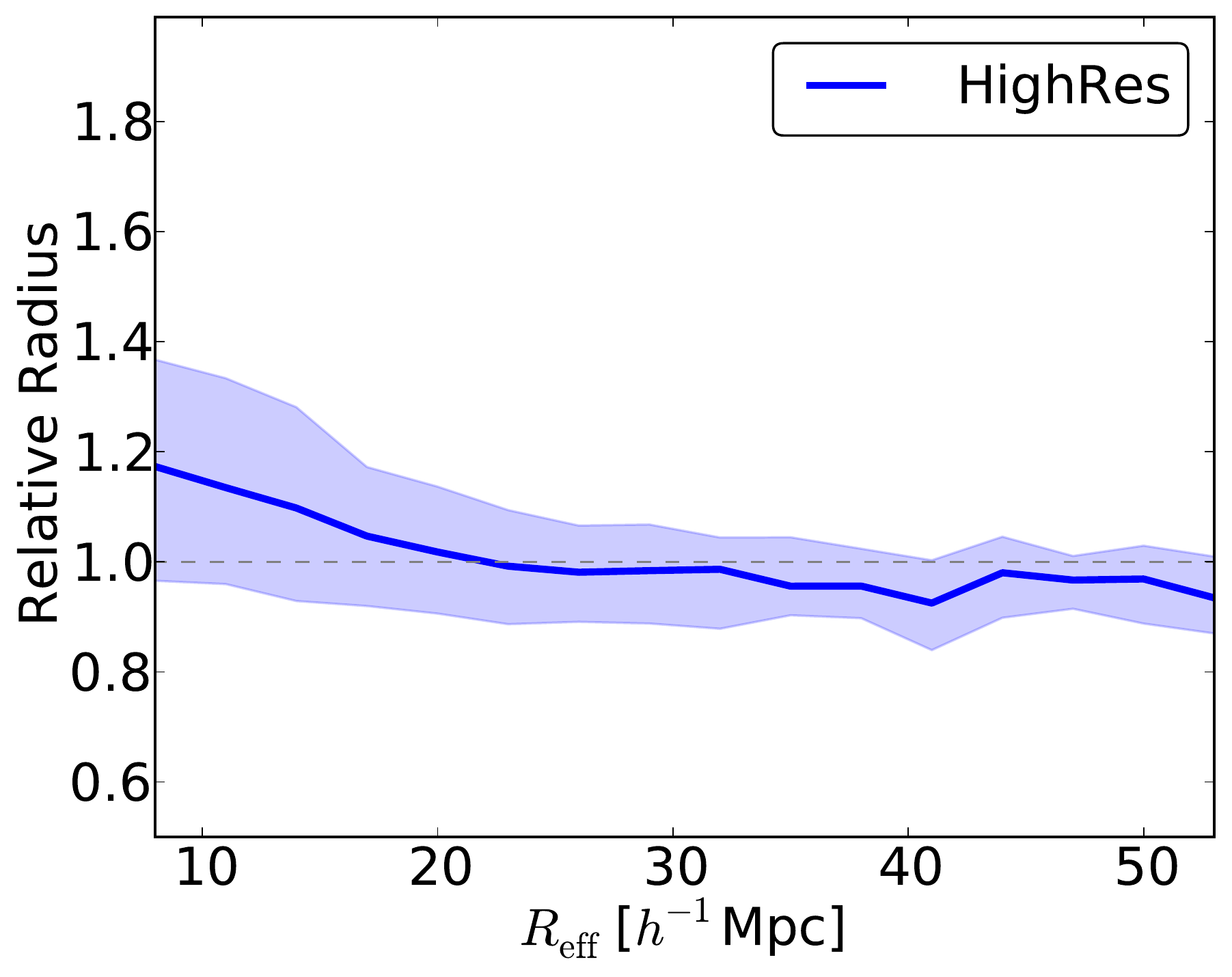}\\
    \includegraphics[width=0.86\columnwidth, angle=0]{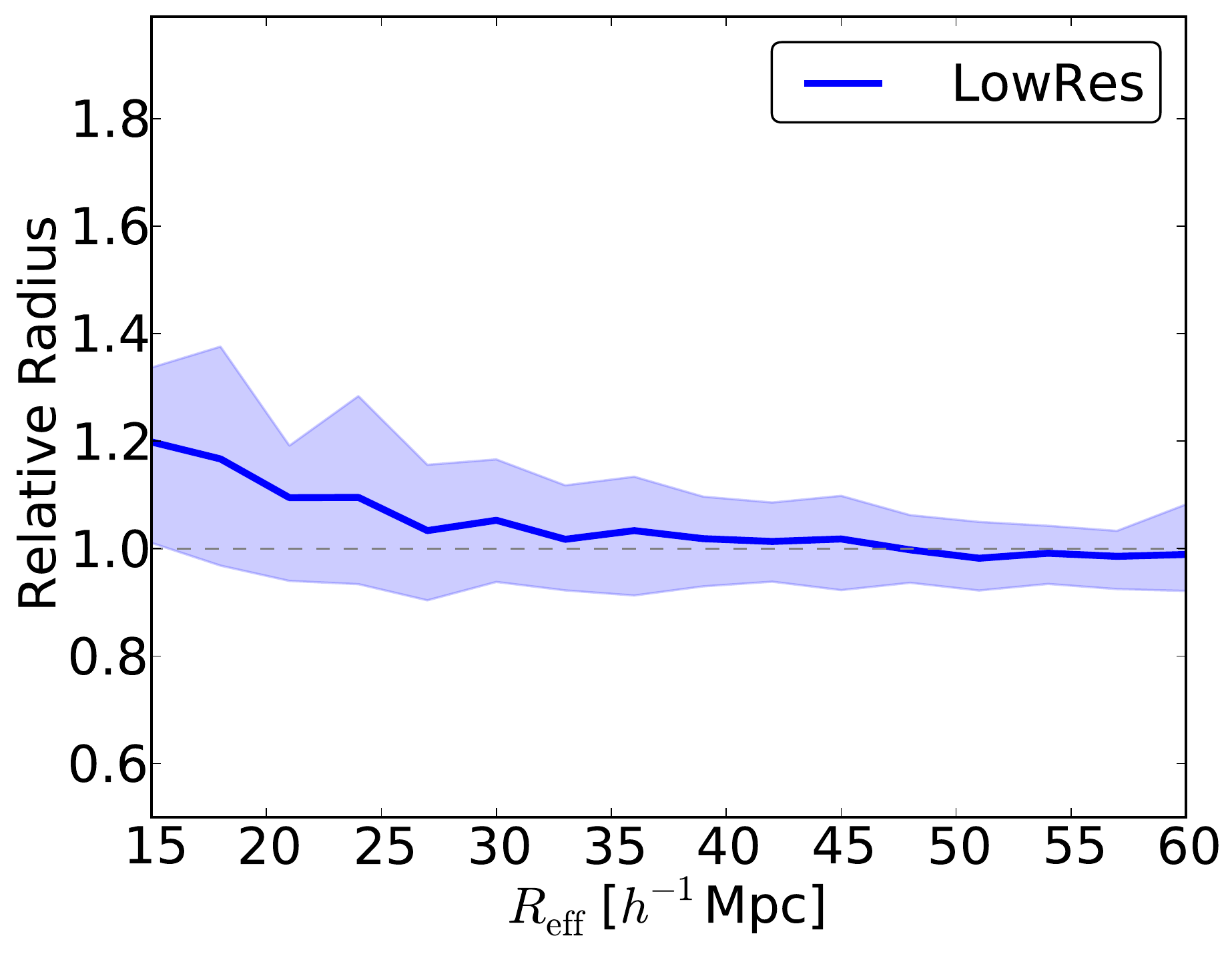}\\  
  \end{tabular}
  \end{center}
 \caption{Relative radius between the peculiar velocity and no-peculiar velocity sample. The shaded regions are the $1\sigma$ Poisson uncertainties. For voids bigger than twice the mean particle separation the radius remains stable both in the \textit{HighRes} sample and in the \textit{LowRes} sample.}
   \label{fig: void relative radius}
\end{figure}

\begin{figure}
\begin{center}
\begin{tabular}{c}
  \includegraphics[width=0.86\columnwidth, angle=0]{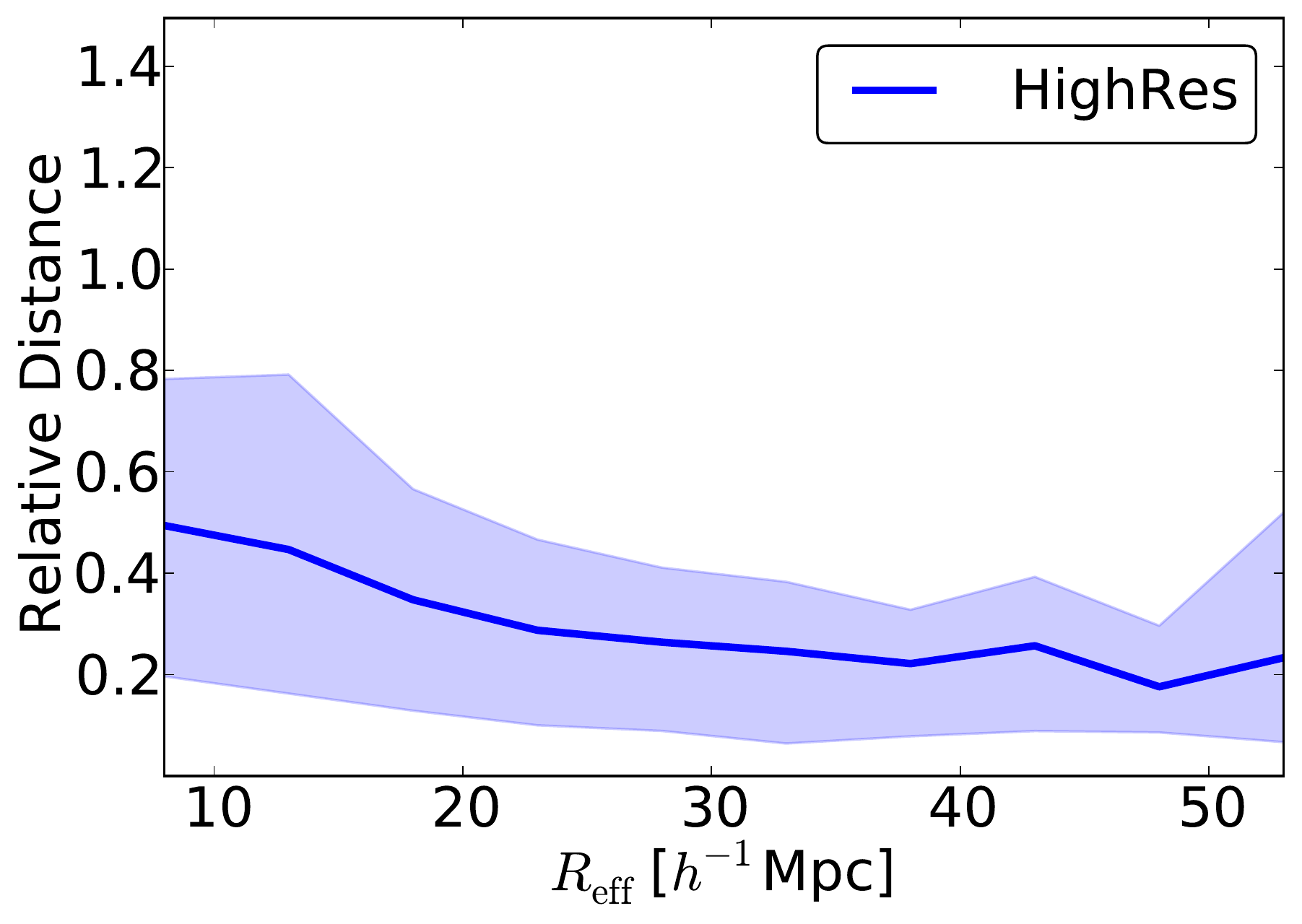}\\
    \includegraphics[width=0.86\columnwidth, angle=0]{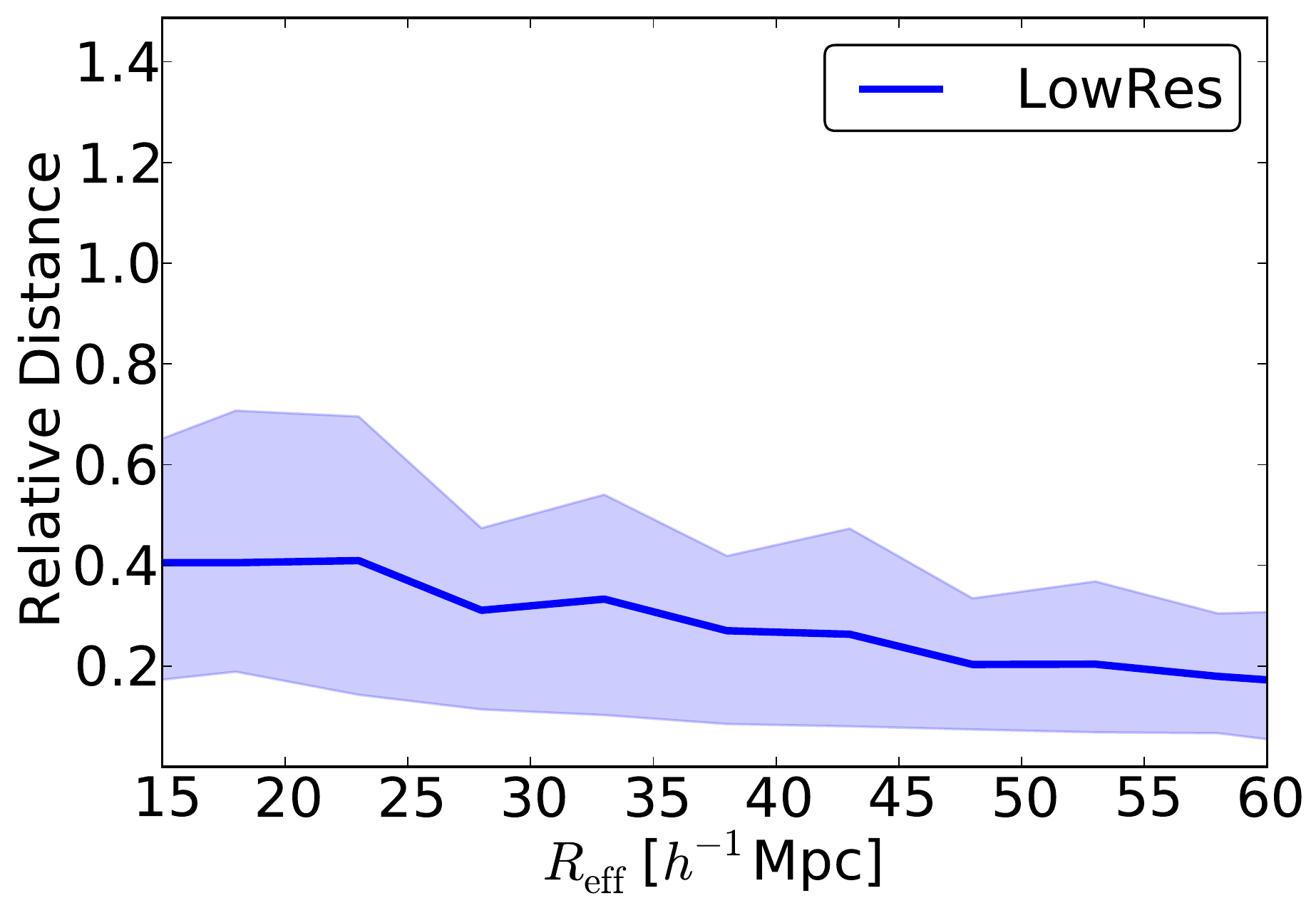}
  \end{tabular}
  \end{center}
 \caption{Relative distance of the position of the macrocenter of voids when adding peculiar velocities. The shaded regions are the $1\sigma$ Poisson uncertainties.}
   \label{fig: void macrocenter}
\end{figure}

\section{Relative Impacts of velocities on void properties}
\subsection{Ellipticity}

For the population of matched voids, we examine the relative change in ellipticity due to velocities by comparing voids in the two catalogues (with and without velocities) for the \textit{HighRes} and \textit{LowRes} samples.
The result of the comparison is shown in Figure \ref{fig: percentage}. 

We find an average reduction of the ellipticity of voids (change lower than 10\% of the value) for both the \textit{HighRes} and \textit{LowRes} samples, although there is significant scatter. For the \textit{HighRes} Sample the ellipticity slightly increases for radii larger than $\sim$ 40 \hmpc, these large voids are not flattened. Recently, \cite{Paz2013} used the redshift space distortion of the void-galaxy cross-correlation function to study the dynamical environment of voids. They found that smaller voids present an inner region in expansion, but their walls are in a collapse stage; while bigger voids are in expansion. Considering this scenario we should expect a flattening of small voids and an enlargement of large voids.

The dynamical effects of the evolution of voids compete with the effect that peculiar velocities will have on how the void finder assigns cells to individual voids. Thus the percentage variation of ellipticity is a result of both these processes, related to the dynamical evolution of voids and to the selection of cells. 

An interesting conclusion is then reached through this work: while voids are generally expanding and one might expect peculiar velocities to elongate individual voids, considering the full cosmic web and the process of finding voids we obtain a different result. Peculiar velocities lead to a thickening of the walls separating voids, thus the void is flattened between growing walls.

The analysis comparing the catalogues with and without peculiar velocities gives us the amount of void ellipticity due to peculiar velocities. The percentage variation is calculated with respect to the ellipticity in redshift space: this point is fundamental, since it allows a direct correction to be applied to measured voids. It means that, if we take voids in redshift space and measure their ellipticity, less than 10\% on average of this ellipticity is due to peculiar velocities --- a correction of the ellipticity taking into account this value enhances the robustness of the Alcock-Paczy\'nski test. Indeed recent applications of the test show that applying a correction for the ellipticity (to take into account the observed constant systematic ellipticity reduction due to peculiar velocities) allows to extract the signal from real data \citep{Sutter2014}.

\subsection{Centers and sizes}

We compute the relative radius between the peculiar velocity and no-peculiar velocity sample (see Figure \ref{fig: void relative radius}). Interestingly, for voids bigger than twice the mean particle separation, the radius remains stable both in the \textit{HighRes} sample and the \textit{LowRes} sample, and for small voids the radii increase by no more than twenty percent. Thus, despite the change in shapes, as discussed above, voids preserve their average volume under distortions from peculiar velocities.

We compare the relative distance of macrocenters, in Figure \ref{fig: void macrocenter}, defined as $d/R_{\rm{eff}}$, where $d$ is the distance between the macrocenters and $R_{\rm{eff}}$ is the effective radius of the void with peculiar velocities. We see that voids are displaced due to the effect of peculiar velocities. Once again the displacement is larger for small voids, but reduces to $\sim 20\%$ for larger voids. However, the change is in the line-of-sight direction, and so uses of the void macrocenter for ISW or lensing analyses are not much affected.

\subsection{Radial density profiles}

\label{sec: density}

\begin{figure*}
\centering
\vspace{-0.3cm}
\begin{tabular}{cccc}

\includegraphics[width=0.53\columnwidth, angle=0]{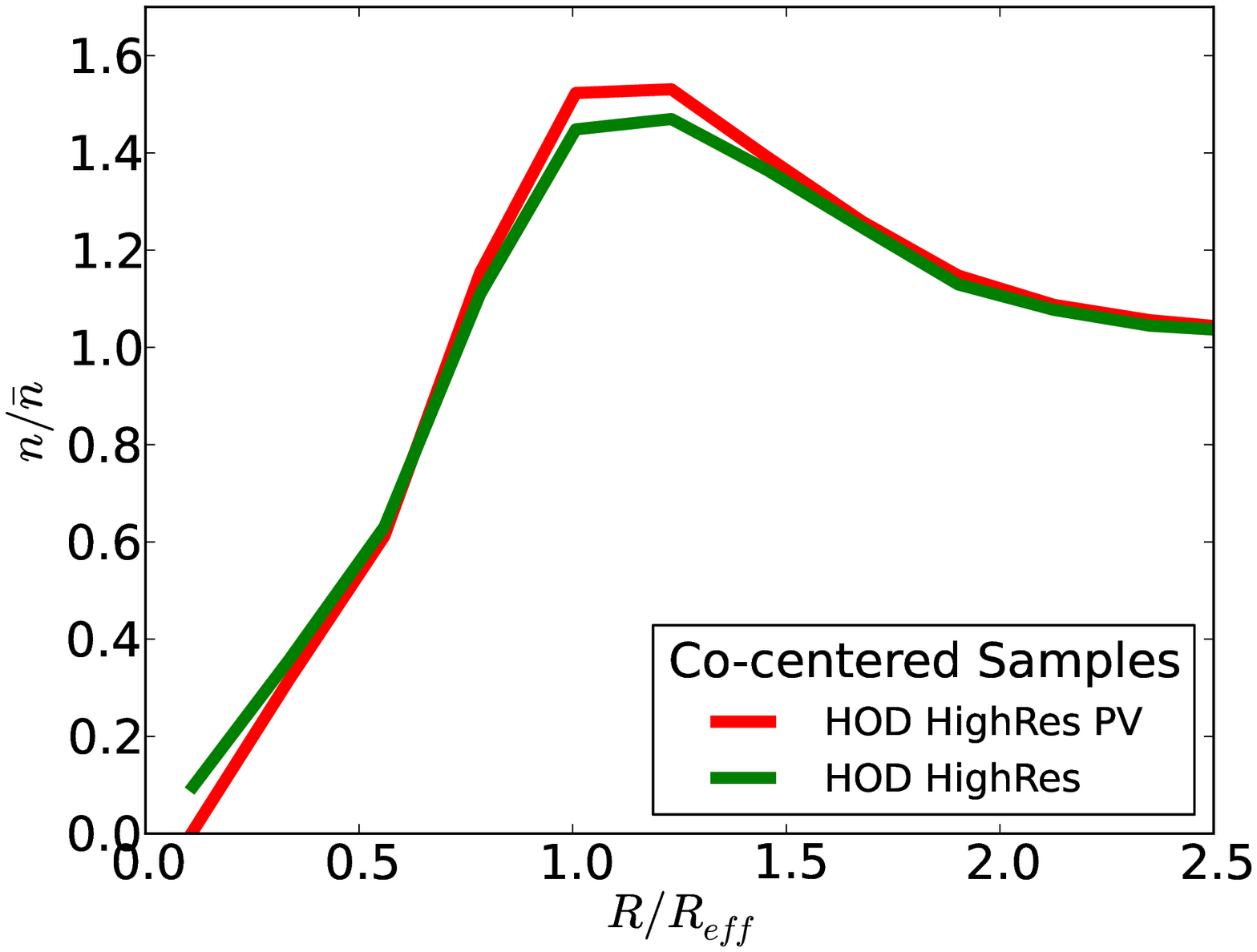}& 
\includegraphics[width=0.53\columnwidth, angle=0]{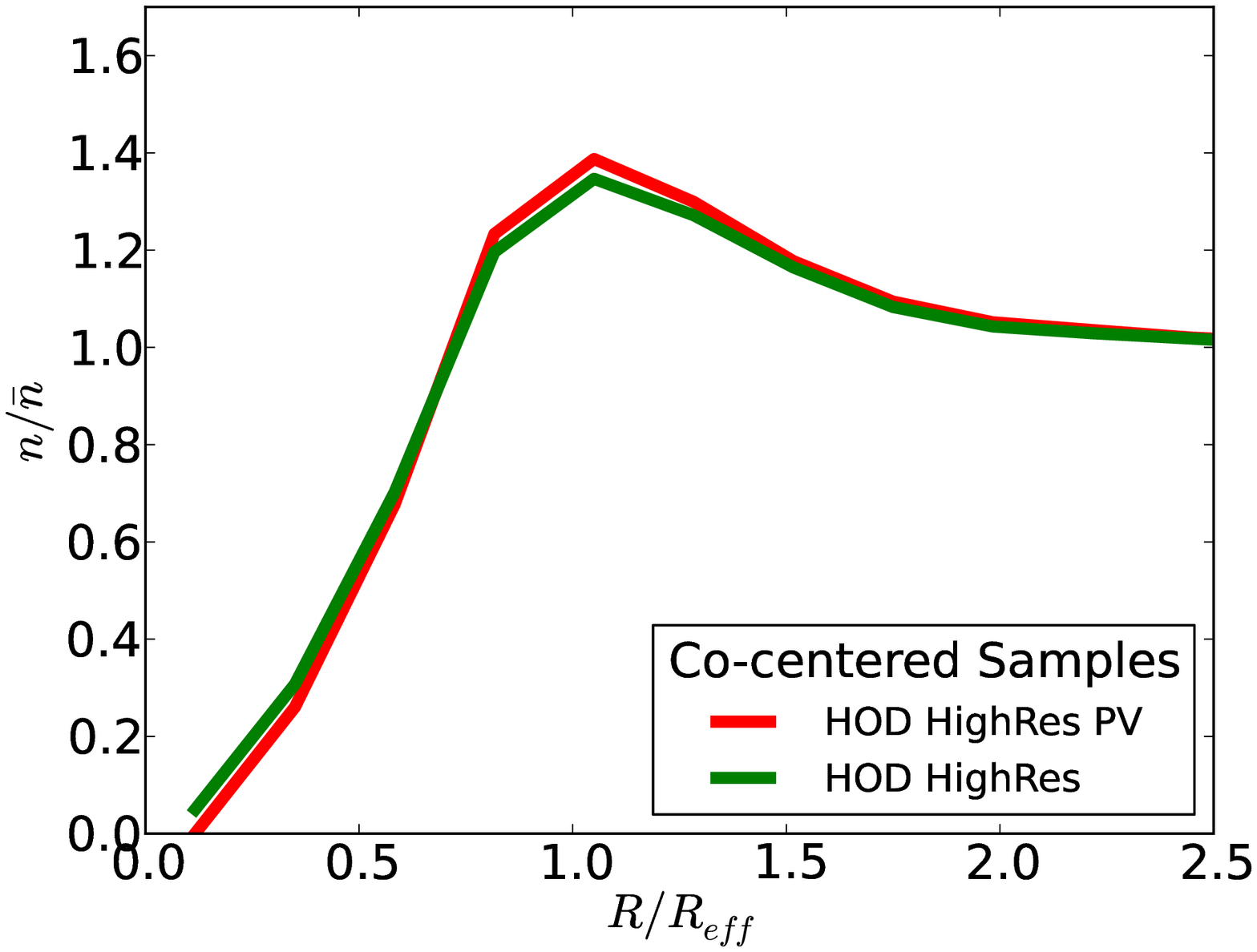}&  
\includegraphics[width=0.53\columnwidth, angle=0]{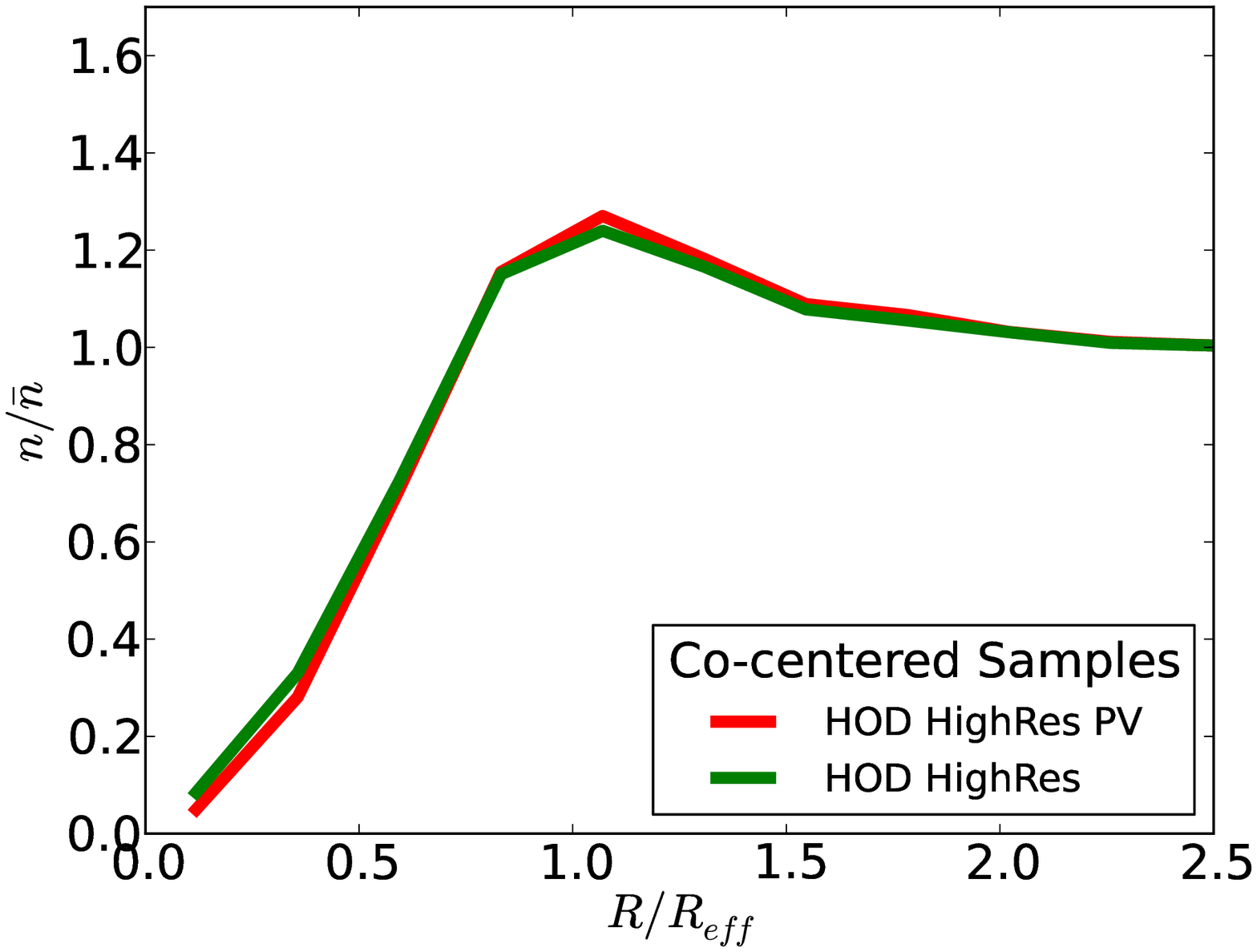}& 
\includegraphics[width=0.53\columnwidth, angle=0]{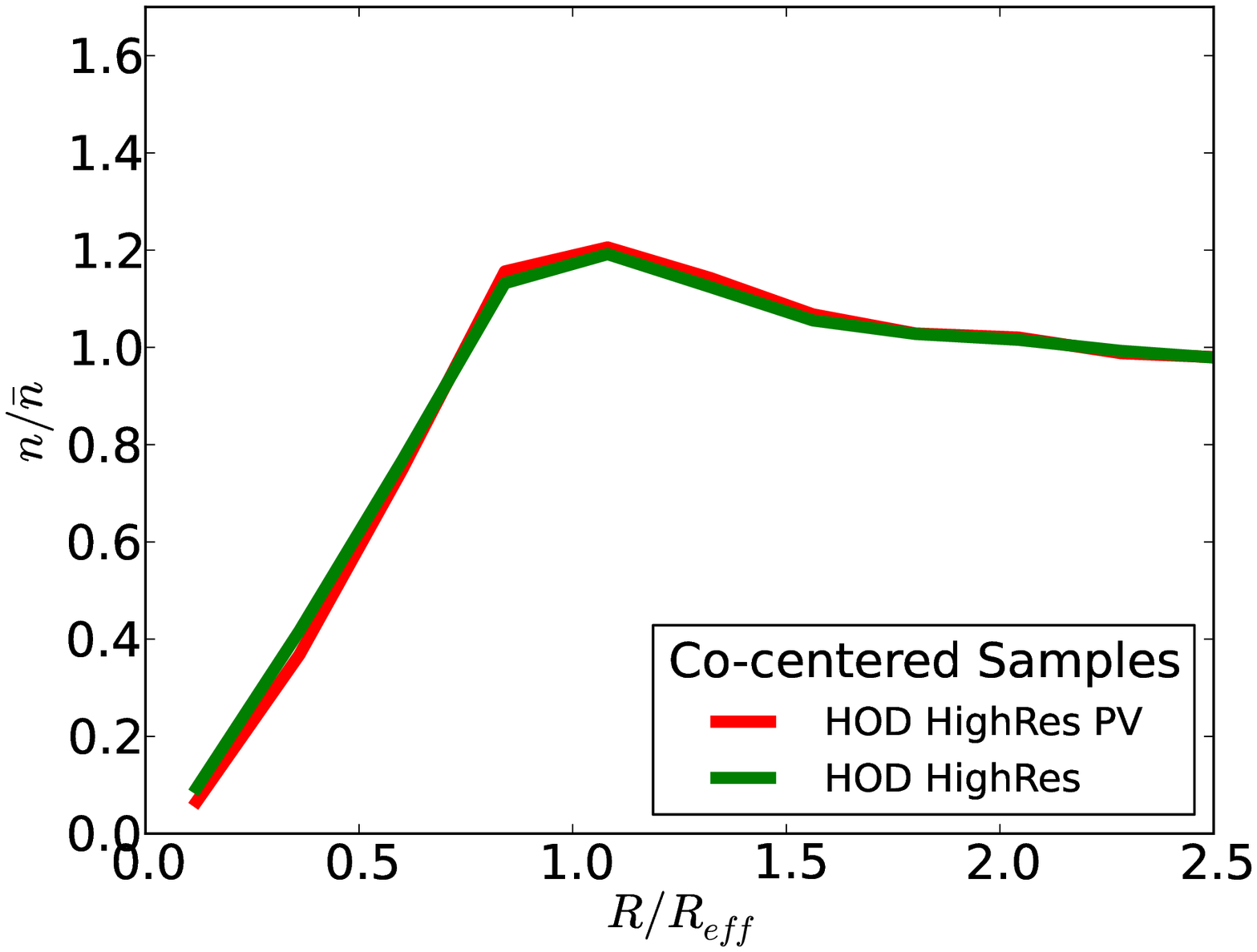}\\
\includegraphics[width=0.53\columnwidth, angle=0]{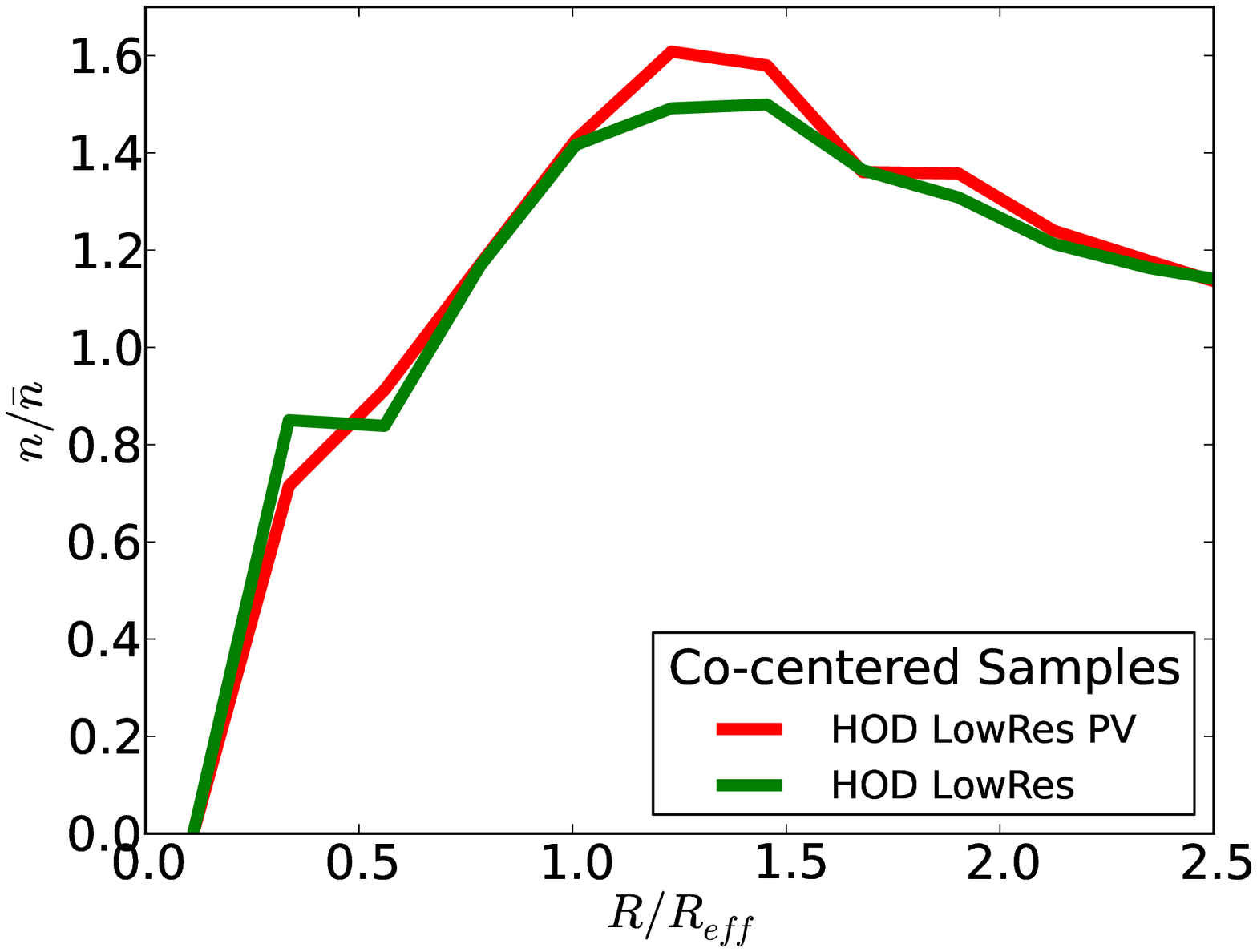}&
\includegraphics[width=0.53\columnwidth, angle=0]{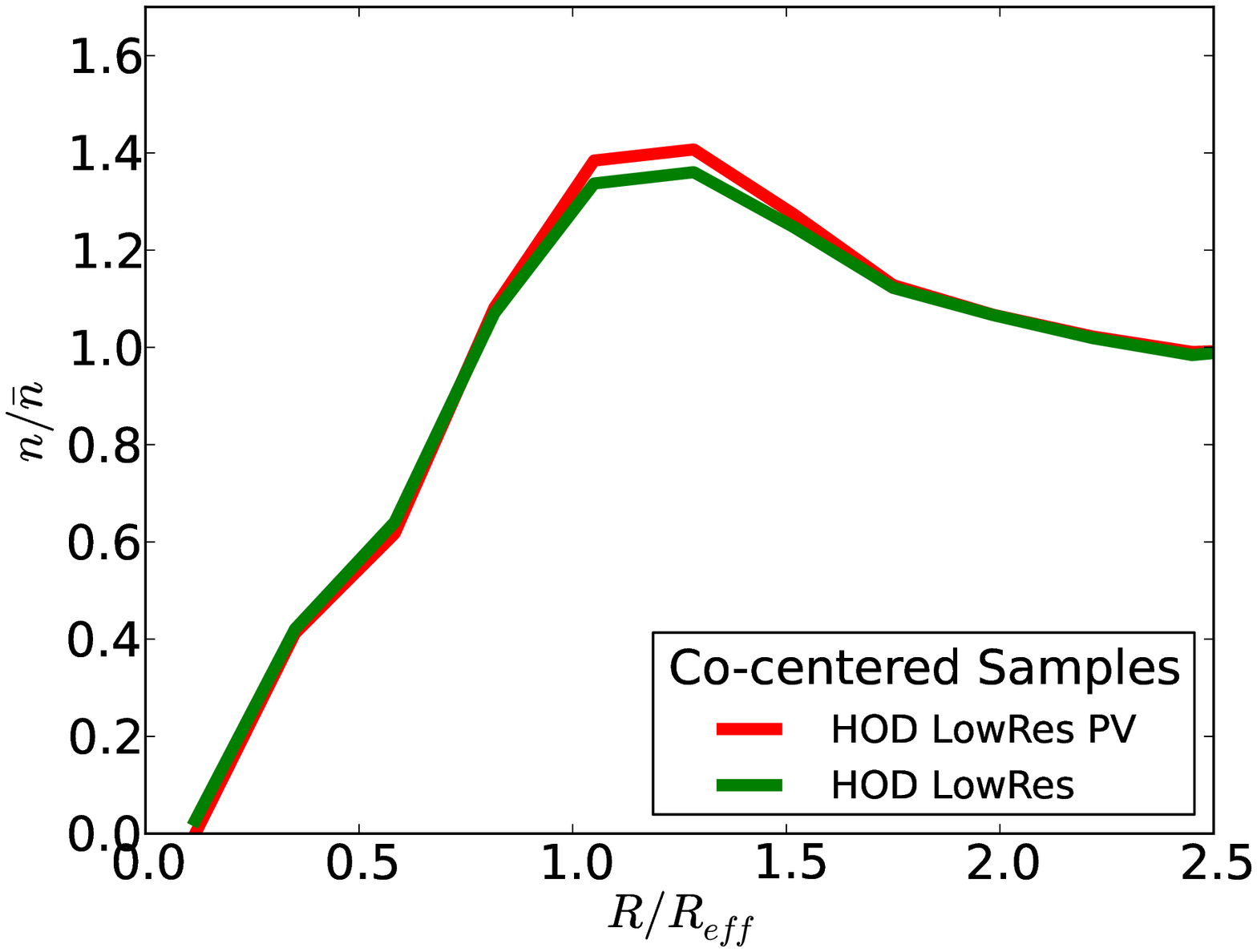}&
\includegraphics[width=0.53\columnwidth, angle=0]{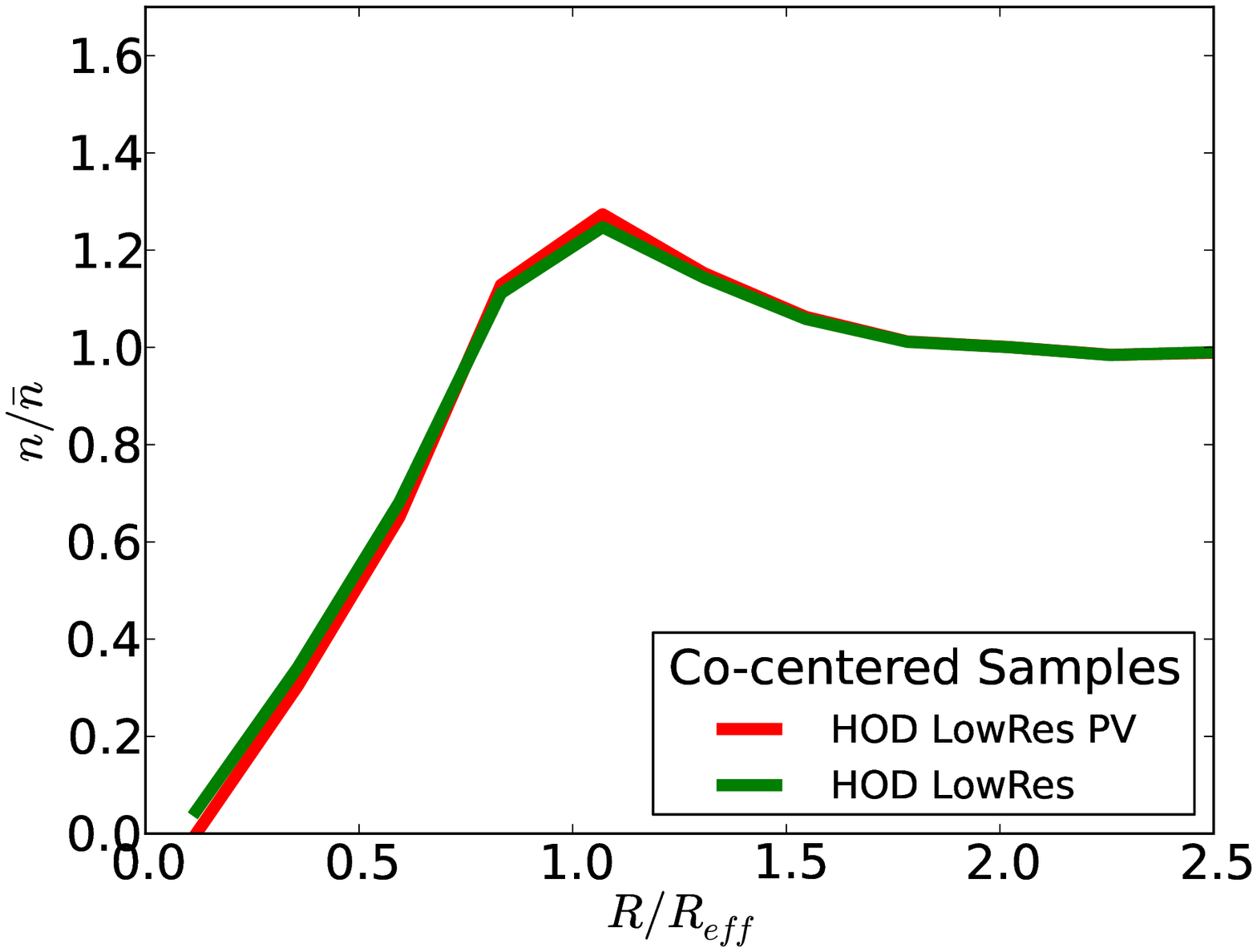}&
\includegraphics[width=0.53\columnwidth, angle=0]{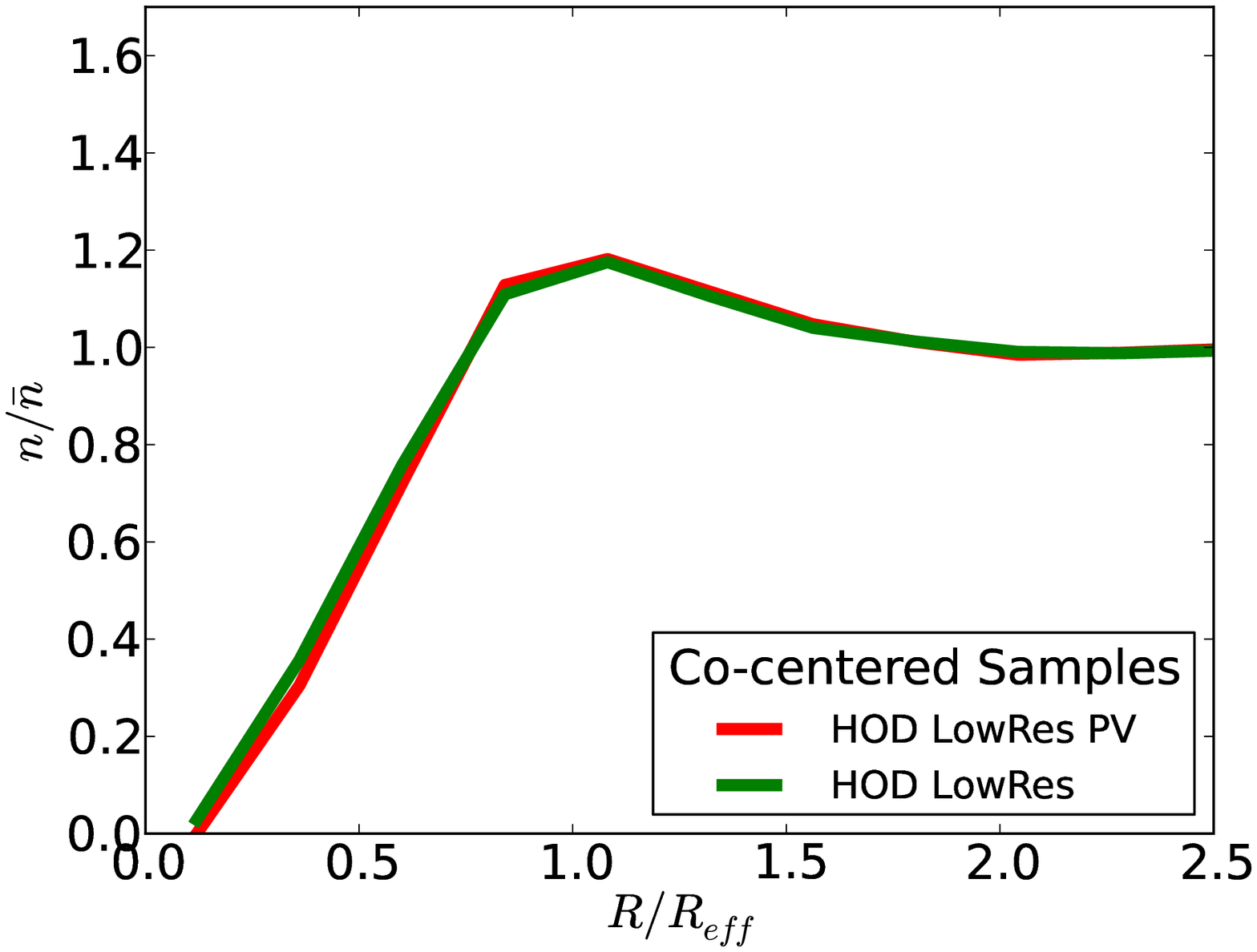}\\          
  \end{tabular}
 \caption{ The figure shows the density profiles of stacks of different sizes for voids with and without peculiar velocities. For both samples, from left to right the stacks are for 15-20, 25-30, 35-40, 45-50 \hmpc radii. Top row is the \textit{HighRes} sample: the density profile is less affected by peculiar velocities while increasing in radius. Bottom row shows the \textit{LowRes} sample: we observe the same trend as for the \textit{HighRes} sample, but the effect of peculiar velocities is washed out at larger radii.}
   \label{fig: void density}
\end{figure*}

We showed in the previous Section that individual small voids can be more affected by peculiar velocities. The stacking of voids can alleviate the effect: even if the shape of the void is not perfectly stable, the averaged density profile can average out fluctuations induced by random perturbations of the tracer particles. This has been suggested in the case of dark matter particle simulations by \cite{Lavaux2012}. We test this claim in a more realistic case using the HOD galaxy samples by comparing the density profiles of stacks in both cases (with and without velocity) for stacks of various radii. Using the technique described in \cite{Sutter2014c}, we consider the co-centered density profiles: we saw that the void macrocenter is slightly displaced due to the effect of velocities. Thus, we shift the centres of voids in the velocity sample to the macrocenter of each matched void and we build the radial profile around the void. 

The profiles presented in Figure \ref{fig: void density} show that the density profiles of stacks are not strongly affected by peculiar velocities, except for the smallest voids. We point out that the use of the co-centering technique is adapted here, since we saw that there is a slight displacement of voids, but the shape and the radii of voids do not change significantly. We find similar results also without the use of the co-centering technique. When using real data, we can just consider the density profiles, without any need to re-center them, because the density profile is not changed but simply displaced. We also notice the slightly enhanced compensation wall for small voids, as described in previous analysis \citep{Sutter2012a, Hamaus2014}. 

As a conclusion, current cosmological constraints relying on density profiles of stacks are only very mildly affected by peculiar velocities; as it is shown in \cite{Sutter2014}, where a constant offset can be used as a first approximate way to take into account their effect. Indeed the profiles are spherically-averaged, so line-of-sight distortions have a reduced impact for the density profile.

\subsection{Abundances}

We compare the abundance of voids in mocks with and without peculiar velocities. We already saw that some voids present in the peculiar velocity catalogue cannot be matched with voids in the catalogue without velocities. Considering only this effect, the number of voids should be higher in the velocity case. But the measure of the void abundance is also sensitive to the voids that are only in the non-peculiar velocity sample but have been disrupted by peculiar velocities --- which means that they are not present in the peculiar velocity catalogue. For most observational probes these voids can be neglected since we will never be able to measure them (absent reconstructions, our measures are always in redshift space). 

\begin{figure}
\begin{center}
\begin{tabular}{c}
  \includegraphics[width=1\columnwidth, angle=0]{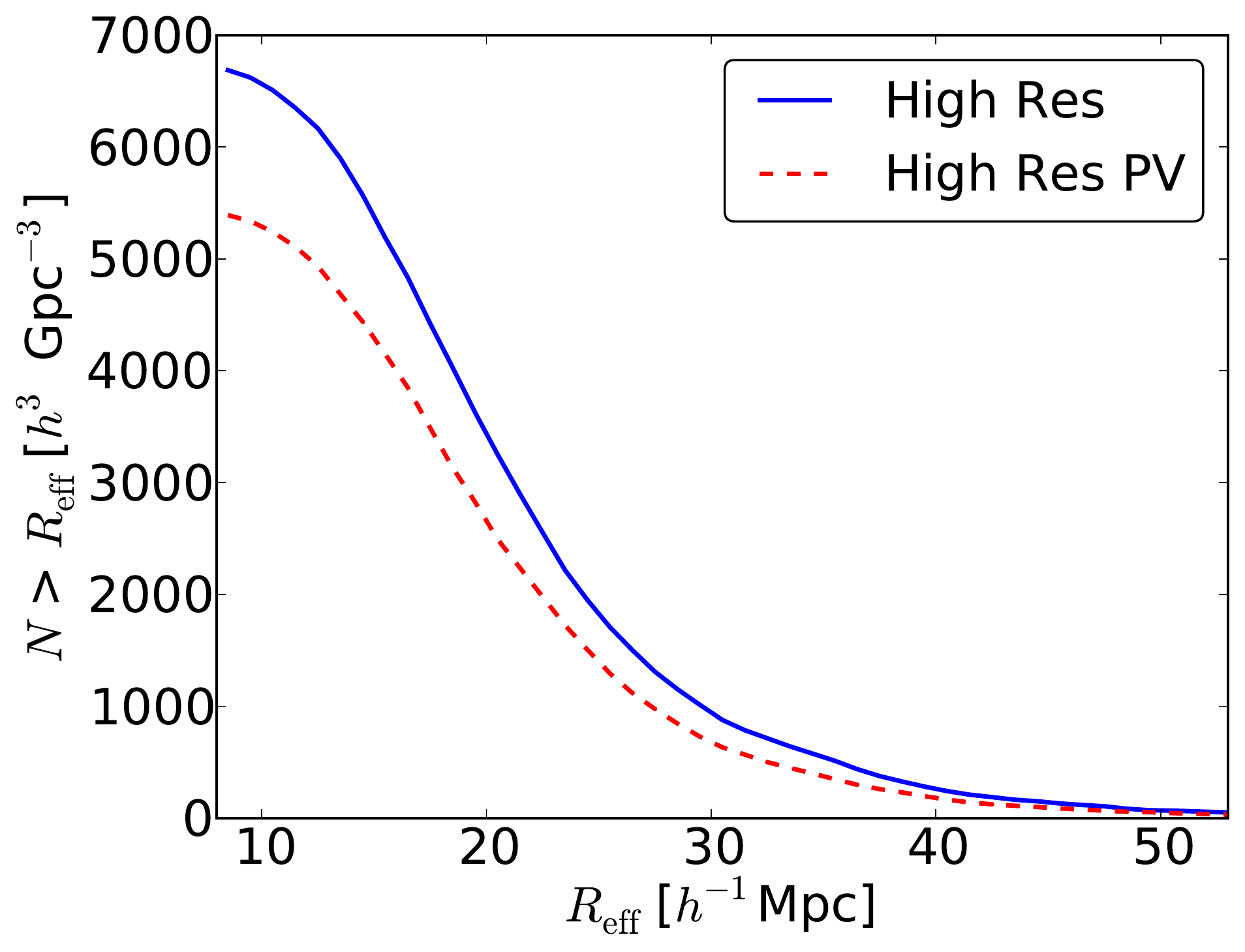}\\
   \includegraphics[width=1\columnwidth, angle=0]{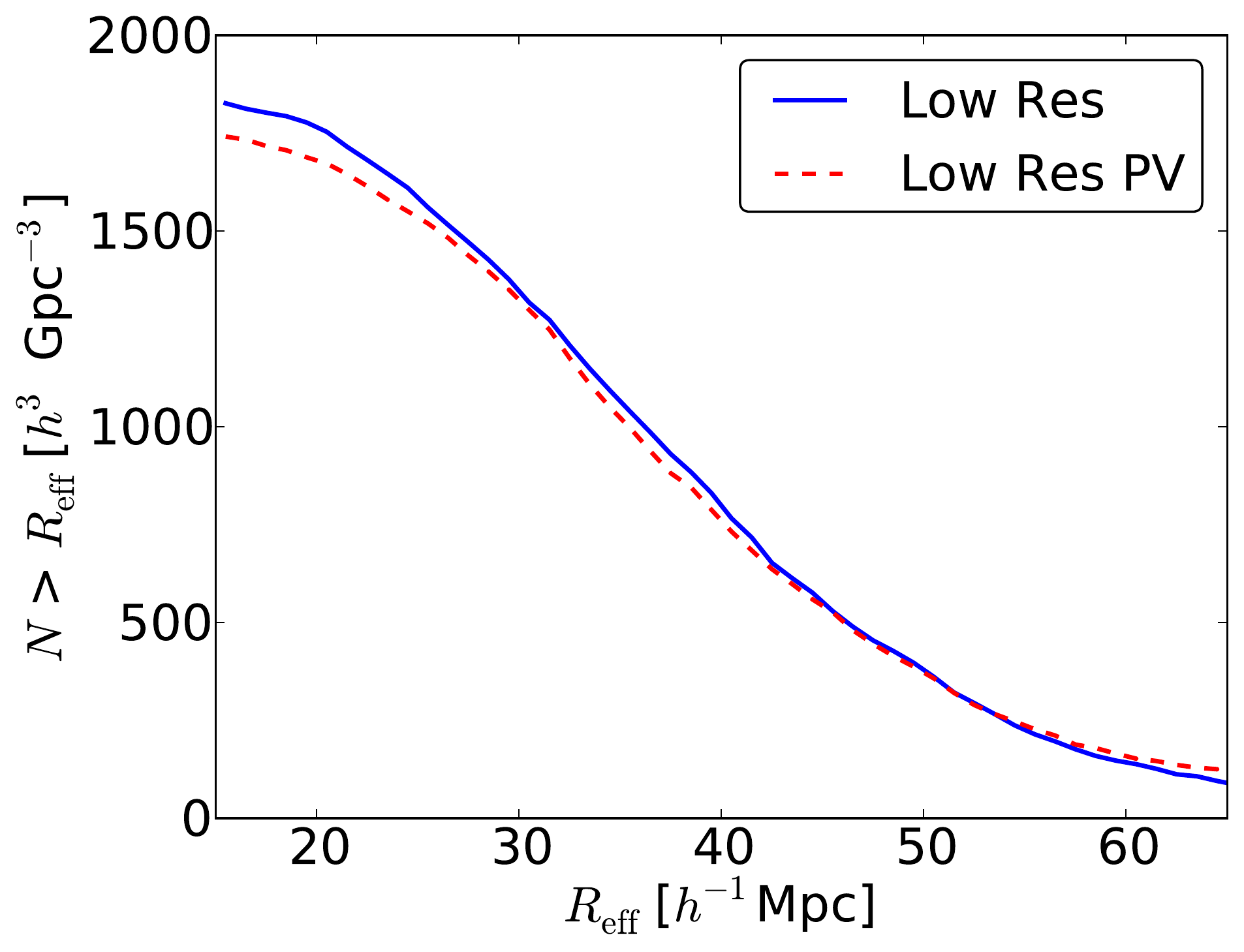}
  \end{tabular}
  \end{center}
 \caption{Void abundances for the \textit{HighRes} and \textit{LowRes} samples, with and without peculiar velocities.}
   \label{fig: void abundances}
\end{figure}

Nevertheless for cosmological applications relying on the comparison of theoretical predictions for void abundances with observations, this effect of peculiar velocities also needs to be considered, since any distortions of the abundance functions due to peculiar velocities must be marginalized over to generate constraints on cosmological parameters. We show in Figure \ref{fig: void abundances} the number of voids for the \textit{HighRes} and \textit{LowRes} samples; with and without peculiar velocities. We observe that peculiar velocities impact the smallest voids: in the \textit{HighRes} sample, the presence of velocities results in a reduction of the abundance of voids. This can be explained by the disruption of small voids by velocities. In the \textit{LowRes} sample this effect is still present, but at a minor level, because with a low density sample the sensitivity to small voids is lower.

In the next Section we comment the application of the considered cuts to real data, and we conclude with general guidelines to master the effect of peculiar velocities in void applications.

\section{Conclusions\label{sec: Discussion}}

With the increase of applications using cosmic voids, the modelling of systematic effects affecting voids becomes crucial. The main systematics one has to deal with when extracting cosmological information from voids is the presence of the peculiar velocities of galaxies. The analysis presented in this paper allows one to tackle the problem of peculiar velocities by studying their effect on void properties through a detailed comparison of two void catalogues obtained from mock galaxy samples with and without the presence of peculiar velocities.

Overall, voids are only mildly affected by peculiar velocities. For most applications that use stacked voids density profiles, abundances, macrocenters, or effective radii (and thus volumes), the effects are negligible. 
In contrast, individual void ellipticities suffer a consistent $\sim 10\%$ reduction (except for the largest voids which suffer an extension).

Additionally, peculiar velocities can strongly disrupt a population of small voids. Thus care must be taken when attempting to include them in an analysis. Voids with radius below twice the mean particle separation and with low density contrast are more likely to be contaminated. Since a population of well-matched voids persists at all scales and density contrast, a trade-off must be considered for cuts based on these properties. Aiming to preserve an acceptable number of voids for the extraction of the signal, it is possible to use the best values determined in Section \ref{sec: matched fraction}. 

We stress that it is certainly possible to include all identified voids in a cosmological analysis, but the effects of peculiar velocities must be accounted for in the subsequent calculations. If instead we wish to 
generate a catalogue of voids with minimal systematic uncertainties, 
the proposed cuts can be directly applied to voids from real data.
For example, applications of the \ap to stacked voids from current data might profit from the exclusion of the smallest voids (for which the cosmological signal is strongly washed out by velocities): at the expense of the statistical weight from small voids, there is a gain in reduced 
shape variance when  limiting the analysis to only larger high-contrast voids.  The cutting methodology discussed in this paper can already be applied within the \vide toolkit.

We have analysed the effect of peculiar velocities on void properties (radius, ellipticity, density contrast, density profiles and abundances) which affect the way we measure voids. 
We have shown that voids are indeed unique places in the cosmic web: they are only slightly affected by 
peculiar velocity distortions, enabling a cleaner examination of any underlying cosmological signal. 

However, in order to achieve sub-percent level precision measurements of cosmological 
parameters, residual systematics must be taken into account. We have shown that the systematic contribution from peculiar velocities is small (and, in many cases, negligible): we considered the application of optimal cuts to reduce the impact of velocities for the extraction of cosmological information from voids in current and future surveys.
The work presented in this paper is a first step towards the understanding and modelling of systematics in a fully self-consistent treatment of cosmic voids as cosmological probes.

\section*{Acknowledgments}

AP thanks Stéphanie Escoffier, Nico Hamaus, and Rien Van de Weygaert for useful discussions. 
AP acknowledges financial support from the grant OMEGA ANR-11-JS56-003-01.
AP and BDW acknowledge support from BDW's Chaire Internationale in Theoretical Cosmology at the Universit\'{e} Pierre et Marie Curie. BDW is partially supported by a senior Excellence Chair of the Agence Nationale de la Recherche (ANR-10-CEXC-004- 01). PMS and BDW acknowledge support from NSF Grant AST-0908902. This research is partially supported by NSF AST 09-08693 ARRA. PMS is supported by the INFN IS PD51 ``Indark''. This work made use of the cluster ``Horizon" at the Institut d’Astrophysique de Paris.

\bibliographystyle{mn2e}
\bibliography{paper}

\bsp

\label{lastpage}

\end{document}